\documentclass{JHEP3}

\usepackage{Pformel,Ppicture,Pmeta}
\usepackage{amsmath,amsfonts,amsthm,times,pst-node}
\usepackage{pstricks,pst-coil}

\newcommand{\Ome}{\tilde{\Omega}_3}
\newcommand{\Omf}{\tilde{\Omega}_4}
\newcommand{\Cnu}{\tilde{C}}
\title{Boundary three-point function on $AdS_2$ D-branes}

\author{Sylvain Ribault
\\
 Laboratoire de Physique Th\'eorique et Astroparticules, UMR5207 CNRS-UM2,
 \\
 Universit\'e Montpellier II, Place E. Bataillon,
 \\
 34095 Montpellier Cedex 05, France 
 \\
 {\footnotesize \tt sylvain.ribault@lpta.univ-montp2.fr }
}

\abstract{ 
Using the $\Hp$-Liouville relation, I explicitly compute the boundary three-point function on $AdS_2$ D-branes in $\Hp$, and check that it exhibits the expected symmetry properties and has the correct geometrical limit. I then find a simple relation between this boundary three-point function and certain fusing matrix elements, which suggests a formal correspondence between the $AdS_2$ D-branes and discrete representations of the symmetry group. Concluding speculations deal with the fuzzy geometry of $AdS_2$ D-branes, strings in the Minkowskian $AdS_3$, and the hypothetical existence of new D-branes in $\Hp$. 
}

\preprint{PTA/07-35 \\ hal-00170145\\ arXiv:0708.3028}

\makeatletter\let\default@color\current@color\makeatother 

\begin{document}

\section{Introduction}

In a recent article \cite{hr06}, Hosomichi and I solved the $\Hp$ model on a disc with boundary conditions corresponding to $AdS_2$ D-branes. 
However the solution was formulated  in terms of variables which are well-adapted to the $\Hp$-Liouville relation, but which obscure the symmetry of the model. For the structure and consequences of the solution to be understood, the symmetry should be made manifest, and this requires some more work. It is particularly important to perform this work in the case of the boundary three-point function because, coming after the bulk three-point function \cite{tes97a} and bulk-boundary two-point function \cite{hr06}, this completes a set of correlation functions from which all others can be obtained. In addition, the boundary three-point function describes the dynamics of boundary condition changing operators, and makes it possible to investigate the structural properties of the model. 

The first purpose of the present article is therefore to explicitly write and analyze the boundary three-point function. This will confirm the correctness of the solution of the $\Hp$ model on the disc. In particular, the geometrical (``minisuperspace'') analysis and the analysis of the symmetries of the boundary three-point function can be understood as further pieces of evidence for the solution proposed in \cite{hr06}. 
The second main purpose of the article is to initiate the study of the structure of the boundary $\Hp$ model, with the eventual aim of confronting it
with general ideas on the structure of boundary conformal field theories. Of course I cannot a priori assume a general result like the relation between fusing matrix and boundary three-point function to hold in the $\Hp$ model, because this non-rational, non-unitary, and non-holomorphically factorizable model violates the assumptions under which such a result is derived. It will however turns out that the boundary three-point function in $\Hp$ can indeed be expressed in terms of certain fusing matrix elements,
provided one introduces a correspondence between the $AdS_2$ D-branes and the discrete representations of the symmetry group, although such representations are absent from the spectrum. 

The calculation of the relevant $\Hp$ fusing matrix elements will not rely on a systematic analysis of the $\Hp$ conformal blocks, which is postponed to future work. Rather, I will make a straightforward and somewhat naive use of the $\Hp$-Liouville relation, which in certain cases yields the $\Hp$ fusing matrix elements in terms of Liouville theory fusing matrix elements. Such an approach is justified a posteriori by the relation with the boundary three-point function. 

The plan of the article is as follows. Section 2 is devoted to defining the boundary three-point function (\ref{defom}) and deriving some features which can be predicted without knowledge of the exact solution, either from a geometrical calculation or from the analysis of the symmetry of the model. In particular, given the symmetry, the three-point function is parametrized by two structure constants $C_\pm$ (\ref{omt}). In section 3, I will use the exact solution \cite{hr06} for checking these predictions, and give an explicit formula (\ref{fef}) for the structure constants. 
Section 4 is devoted to the computation of fusing matrix elements in $\Hp$, and to their relation (\ref{chfh}) with the boundary three-point function. This will require the formal introduction of discrete representations. The concluding section 5 will offer some speculations which are inspired by these results.

This article can be thought of as a follow-up of \cite{hr06}, which is briefly summarized in \cite{hos07}. Nevertheless, the necessary results on the $\Hp$ model on a disc \cite{pst01,hr06} will be recalled, although not explained in detail. The necessary results on Liouville theory, which come from the works \cite{fzz00,tes00,pt01,tes01a}, will also be recalled, mostly in the conventions of the short review \cite{pon03}. 

\zeq\section{The three-point function: predictions}

\subsection{Geometrical description \label{submini}}

The aim of this subsection is to predict the geometrical limit of the boundary three-point function in $\Hp$. I will first recall (from \cite{pst01}) which model is obtained as the geometrical limit of the $\Hp$ model, and which quantities should have well-defined limits. This will lead to the definition of a geometrical three-point function, which will then be explicitly computed.

\paragraph{Geometry of $\Hp$ and of the $AdS_2$ D-branes.} 
 
The three-dimensional Euclidean space $\Hp$ can be defined as the set of two-by-two Hermitian matrices $h$ of determinant one, and parametrized by three coordinates $(\phi,\gamma,\bg)$ such that $h=\left( \begin{smallmatrix} e^\phi &\ &e^\phi\bg \\ e^\phi\g &\ &e^\phi \g\bg+e^{-\phi} \end{smallmatrix}\right)$. The space $\Hp$ can also be seen as the right coset $\SLC/SU(2)$, on which an $\SLC$ symmetry group acts by left multiplication; the resulting action of $g\in\SLC$ on the Hermitian matrix $h$ is $g\cdot h = ghg^\dagger$.
The D-branes of interest are Euclidean $AdS_2$ branes, which should more accurately be called $H_2^+$ branes. They are defined by equations of the type $\Tr \Omega h = 2\sinh r$ where the real parameter $r$ determines the curvature of $H_2^+$ while the Hermitian matrix $\Omega$ determines its orientation. Such a D-brane intersects the $\phi=\infty$ boundary of $\Hp$, which is a two-sphere $S^2$, and the intersection is a great circle, with an equation of the type $|\g - \g_0| = R_0$ or $ \Re (\mu_0 \g) = \lambda_0 $. 

Let me fix the orientation of the $AdS_2$ branes, and consider only D-branes with the same matrix $\Omega=\left(\begin{smallmatrix} 0 & 1\\ 1 & 0 \end{smallmatrix}\right)$, the same great circle at infinity $\g+\bg=0$, and the same preserved $\SLR$ subgroup $\left\{g=\left(\begin{smallmatrix} a & ic \\ -ib & d \end{smallmatrix}\right),\ ad-bc=1,\ a,b,c,d\in \R\right\}$ of the $\SLC$ symmetry group. 
This assumption ensures that the theory of open strings stretched between two such D-branes enjoys a maximal amount of symmetry. A further assumption is needed for the theory of open strings on $AdS_2$ branes to have a geometrical description: open strings should reduce to point particles, which is only possible if they have both ends on the same D-brane. In this subsection I will therefore assume all involved $AdS_2$ branes to have the same parameter $r$, thus the same equation $e^\phi(\g+\bg)=2\sinh r$. The theory of open strings on this D-brane then has a well-defined geometrical description in the {\it minisuperspace limit}, as the quantum mechanics of a point particle in $AdS_2$. 

\paragraph{Point particles in $AdS_2$.} 

Point particles in the Euclidean $AdS_2$ are described by their wavefunctions: complex-valued functions on $AdS_2$. Their spectrum, namely the space of such functions, can be organized according to the action of the $\SLR$ symmetry group. Namely, the spectrum is generated by functions
\bea
\Psi^\ell(t|h) = \left( |\g+it|^2 e^\phi +e^{-\phi}\right)^\ell\ ,
\eea
which belong to continuous representations of $\SLR$ of spins
$\ell\in -\frac12+i\R$ and Casimir eigenvalues $-\ell(\ell+1)$, and $t\in \R$ is the isospin variable. The transformation of such functions under the action of $g\in \SLR$ is indeed
\bea
\Psi^\ell(t|g\cdot h) = |ct-d|^{2\ell} \Psi^\ell(g\cdot t|h)\ ,\qquad
g\cdot t=\frac{at-b}{-ct+d}\ .
\label{gpl}
\eea
Let me define the geometrical three-point function on an $AdS_2$ brane of parameter $r$ as 
\bea
\Omega_3^{geom} \equiv \int dh\ \delta(e^\phi(\g+\bg)-2\sinh r)\ \prod_{i=1}^3\Psi^{\ell_i}(t_i|h)\ ,
\eea
where $dh=e^{2\phi}d\phi\ d^2\g $ is the $\SLC$-invariant measure on $\Hp$. The purpose of this subsection is to obtain the explicit expression of $\Omega_3^{geom}$. 

\paragraph{Calculation of $\Omega_3^{geom}$.}

The calculation goes as follows (neglecting numerical factors). Perform the integral over $\g+\bg$ and write $\gamma = e^{-\phi}\sinh r - i\rho $ with $\rho\in\R$, then perform the shift $\phi \rar \phi +\log \cosh r$. This yields
\bea
\Omega_3^{geom} = (\cosh r)^{\sum \ell_i +2} \int e^{\phi} d\phi\ d\rho\ \prod_{i=1}^3 \left(|\rho-t_i|^2 e^\phi +e^{-\phi}\right)^{\ell_i} \ .
\eea
Having made the $r$-dependence explicit, the next step is to make the $t_i$-dependence explicit:
\bea
\Omega_3^{geom} = (\cosh r)^{\sum \ell_i +2}\ |t_{12}|^{\ell_{12}^3}|t_{13}|^{\ell_{13}^2}|t_{23}|^{\ell_{23}^1}\ C^{geom}(\ell_1,\ell_2,\ell_3)\ ,
\label{otg}
\eea 
with the notations $t_{12}=t_1-t_2$ and $\ell_{12}^3=\ell_1+\ell_2-\ell_3$. This formula can be derived by using the $\SLR$ symmetry of $\Omega_3^{geom}$, and its explicit expression
in the limit $t_3\rar \infty$, after performing the change of variables $(\phi,\rho)\rar (\phi-\log|t_{12}|,t_{21}\rho+t_1)$. This also provides
the integral expression of $C^{geom}$, the geometrical limit of the three-point structure constant at $r=0$:
\bea
C^{geom} = \int e^\phi d\phi\ d\rho\ \left(\rho^2 e^\phi +e^{-\phi}\right)^{\ell_1} \left((\rho-1)^2 e^{\phi} +e^{-\phi}\right)^{\ell_2} e^{\ell_3\phi}\ .
\eea
Now introduce variables $(x_1,x_2)=(e^\phi \rho, e^\phi(1-\rho))$, while allowing $e^\phi$ to take all real values,
\bea
C^{geom} = \int_{\R^2} dx_1\ dx_2\ |x_1+x_2|^{-\ell_{12}^3-1} (1+x_1^2)^{\ell_1} (1+x_2^2)^{\ell_2}\ .
\eea
Inserting $1=\int dy\ \delta(y+x_1+x_2) $ and $\delta(y+x_1+x_2) = \int d\theta\ e^{i\theta (y+x_1+x_2)}$ yields
\bea
C^{geom} &=& \int d\theta \int dy\ dx_1\ dx_2\ e^{i\theta (y+x_1+x_2)} |y|^{-\ell_{12}^3-1} (1+x_1^2)^{\ell_1} (1+x_2^2)^{\ell_2}
\label{cgp}
\\
&=& 2^{\ell_1+\ell_2} \frac{\G(-\ell_{12}^3)\sin\frac{\pi}{2} \ell_{12}^3}{\G(-\ell_1)\G(-\ell_2)} \int_0^\infty d\theta\ \ \theta^{-\ell_3-1} K_{-\ell_1-\frac12}(\theta) K_{-\ell_2-\frac12}(\theta)\ ,
\eea
\bea 
 \boxed{C^{geom} = 
\G(-\tfrac12(\ell_{123}+1)) \frac{\G(-\frac12\ell_{12}^3)\G(-\frac12\ell_{13}^2) \G(-\frac12\ell_{23}^1)}{ \G(-\ell_1)\G(-\ell_2)\G(-\ell_3) }} \ ,
\label{cgeom}
\eea
% Gradshteyn - Ryzhik 8.432.5 p. 907 (integral leading to Bessel function K)
% ibid                6.576.4 p. 676 (integral of two Bessel functions and a power)
where I used standard formulas \cite{gr65} for the Bessel function with imaginary argument $K$, and the integral formula (\ref{iyp}). (And a new notation: $\ell_{123}=\ell_1+\ell_2+\ell_3$.)

The formula for $C^{geom}$ is permutation-symmetric, which is a basic check of its correctness. It vanishes for discrete spins $\ell \in \N$, which explains the absence of discrete representations in the spectrum, in spite of their appearance in tensor products of continuous representations. And it will be shown to agree with the geometrical limit of the exact open string three-point function in subsection \ref{subgeom}.

\subsection{Symmetry}

Let me leave the geometrical limit and consider more general boundary three-point functions, where open strings can have their ends on different $AdS_2$ D-branes. I will now derive the constraints on the boundary three-point function which follow from the assumed symmetries of the model. The symmetry group of the model is an infinite-dimensional loop group, whose Lie algebra is the affine Lie algebra $\asl$. The three-dimensional horizontal subgroup will be most relevant in the following. 

\paragraph{Action of the symmetry on the open strings.} 

The global structure of the horizontal subgroup of the symmetry group of the $\Hp$ model on the disc
was understood only recently \cite{hr06}, because it differs from the $\SLR$ group which is present in the geometrical limit, and which had naively been expected to be present in the general case as well. 
The correct symmetry group is actually $\USLR$, the universal covering group, whose elements are pairs $(g,[T])$ with $g=\left(\begin{smallmatrix} a & ic \\ -ib & d \end{smallmatrix}\right)$ an element of the same $\SLR$ subgroup of $\SLC$ as before, and $[T]\in \Z$ an integer. The group multiplication law is $(g,[T])\cdot (g',[T']) = (gg',[T]+[T']+[g,g'])$ where $[g,g']\in\{0,1\}$ is the integer part of $T(g)+T(g')$, with $T(g)\in [0,1[$ a timelike coordinate on $\SLR$. (The elements of the additive group $\R$ can similarly be viewed as pairs of an element of $[0,1[$ and an integer, whose addition law would then be similar to the present $\USLR$ multiplication law.)
The action of $\USLR$ on vertex operators is\footnote{The present convention for the
sign of the exponent differs from \cite{hr06}. The present convention will be consistent with the chosen conventions in Liouville theory through the $\Hp$-Liouville relation. I believe that the conventions in \cite{hr06} were not consistent in this respect.}
\bea
(g,[T])\cdot {}_r\Psi^{\ell}(t|w)_{r'} = |ct-d|^{2\ell} e^{-(k-2)(r-r')\left([T]+\frac12+\frac12 \sgn(t-\frac{d}{c})\right)} {}_r\Psi^{\ell}(g\cdot t|w)_{r'}\ ,
\label{gtp}
\eea
where the vertex operator ${}_r\Psi^{\ell}(t|w)_{r'}$, whose position on the boundary of the worldsheet is $w\in \R$, describes an open string stretched between two $AdS_2$ branes with the same orientation and parameters $r$ and $r'$; and $k>2$ is the level of the $\Hp$ model, which is related to the central charge by $c=\frac{3k}{k-2}$, and will sometimes be replaced with the equivalent parameter $b^2=\frac{1}{k-2}$. Like in the geometrical limit, the spectrum is purely continuous with spins $\ell\in -\frac12+i\R$. 

\paragraph{Definition of the boundary three-point function.}

The boundary three-point function is defined as the expectation value
\bea
\Omega_3 = \la {}_{r_{31}} \Psi^{\ell_1}(t_1|w_1) _{r_{12}} \Psi^{\ell_2}(t_2|w_2)_{r_{23}} \Psi^{\ell_3}(t_3|w_3)_{r_{31}}\ra \ .
\label{defom}
\eea
From the point of view of two-dimensional conformal field theory, this describes the insertion of three vertex operators on the circular boundary of a disc worldsheet. From the target space point of view, this describes three open strings stretched between three $AdS_2$ branes of parameters $r_{12},r_{23},r_{13}$, whose identical orientation means they coincide at infinity. (For convenience, only two dimensions of $\Hp$ are represented here, and the sphere $S^2$ at infinity is represented as a dashed circle. The open string states are represented as well-localized wiggly lines, although in reality the operators $\Psi^{\ell_i}$ rather correspond to momentum eigenstates.) 
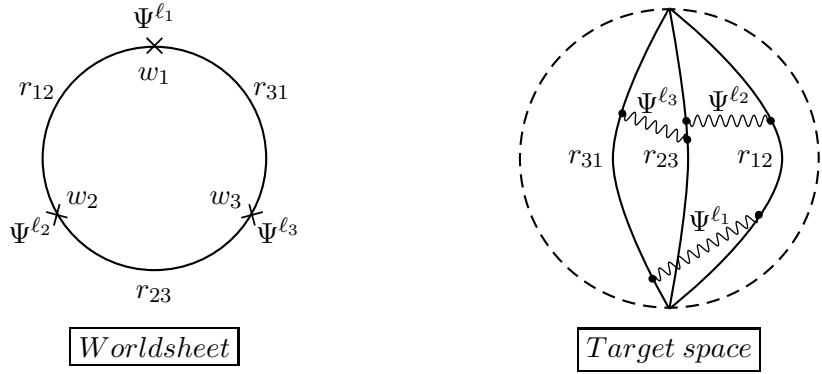
\begin{center}
\begin{tabular}{ccc}
 \psset{unit=.5cm}
 \pspicture[](-4,-4.5)(4,4)
 \pscircle(0,0){3}
 \psdots[dotstyle=+,dotangle=45,dotscale=2](0,3)
 \rput{*0}(0,3.8){$\Psi^{\ell_1}$} \rput{*0}(0,2.2){$w_1$}
 \rput{-120}{
 \psdots[dotstyle=+,dotangle=45,dotscale=2](0,3)
 \rput{*0}(0,3.8){$\Psi^{\ell_3}$} \rput{*0}(0,2.2){$w_3$} 
 }
 \rput{120}{
 \psdots[dotstyle=+,dotangle=45,dotscale=2](0,3)
 \rput{*0}(0,3.8){$\Psi^{\ell_2}$} \rput{*0}(0,2.2){$w_2$} 
 }
 \rput{-60}{\rput{*0}(0,3.6){$r_{31}$}}
 \rput{60}{\rput{*0}(0,3.6){$r_{12}$}}
 \rput(0,-3.6){$r_{23}$}
 \endpspicture
 & \hspace{2cm} &
 \psset{unit=.5cm}
 \pspicture[](-4,-4)(4,4)
 \pscircle[linestyle=dashed](0,0){4}
 \pscurve(0,4)(3,0)(0,-4)
 \pscurve(0,4)(.5,0)(0,-4)
 \pscurve(0,4)(-1.5,0)(0,-4)
 \rput(2.3,0){$r_{12}$}
 \rput(-.2,0){$r_{23}$}
 \rput(-2.2,0){$r_{31}$}
 \pscoil[coilarm=.1,coilwidth=.3,coilaspect=0,linewidth=.5pt]{*-*}(.45,1)(2.7,1)
 \pscoil[coilarm=.1,coilwidth=.3,coilaspect=0,linewidth=.5pt]{*-*}(-1.25,1.2)(.48,.5)
 \pscoil[coilarm=.1,coilwidth=.3,coilaspect=0,linewidth=.5pt]{*-*}(-.45,-3.2)(2.38,-1.5)
 \rput(1.6,1.6){$\Psi^{\ell_2}$}
 \rput(-.3,1.55){$\Psi^{\ell_3}$}
 \rput(1.1,-1.6){$\Psi^{\ell_1}$}
 \endpspicture
 \\ \boxed{Worldsheet} & & \boxed{Target\ space} 
\end{tabular}
\end{center}
The dependence of the three-point function on the boundary coordinates $w_i\in \R$ is determined by conformal symmetry to be a factor $|w_{12}|^{\Delta_{\ell_3} -\Delta_{\ell_1}-\Delta_{\ell_2}} |w_{23}|^{\Delta_{\ell_1} -\Delta_{\ell_2}-\Delta_{\ell_3}} |w_{13}|^{\Delta_{\ell_2}-\Delta_{\ell_1} -\Delta_{\ell_3}}$, which will be omitted henceforth. Here $\Delta_\ell = -\frac{\ell(\ell+1)}{k-2}$ is the conformal weight of $\Psi^\ell$, and $w_{12}=w_1-w_2$. It is however necessary to keep track of the order of the fields on the boundary of the disc. The three-point function is indeed expected to be invariant under cyclic permutations, but not under a permutation of two fields. This differs from the full permutation symmetry of the boundary three-point function of say Liouville theory. This is because the $\Hp$ boundary field ${}_r\Psi^{\ell}(t|w)_{r'}$ and its symmetry transformation (\ref{gtp}) are nontrivially affected by the exchange of the two boundary conditions $r,r'$. In other words, the boundary theory is not invariant under worldsheet parity. Here I am assuming the boundary to be oriented counterclockwise, and the boundary operators to come in the order $1,2,3$ like in formula (\ref{defom}). 

\paragraph{Solving the $\USLR$ symmetry condition.}

The $\USLR$ symmetry condition on the boundary three-point function is 
\bea
\la (g,[T])\cdot \Psi^{\ell_1}\ (g,[T])\cdot \Psi^{\ell_2}\ (g,[T])\cdot \Psi^{\ell_3} \ra = \la \Psi^{\ell_1}\ \Psi^{\ell_2}\ \Psi^{\ell_3} \ra\ ,
\eea
which explicitly reads
\begin{multline}
 \Omega_3\left(\frac{at_1-b}{-ct_1+d},\frac{at_2-b}{-ct_2+d},\frac{at_3-b}{-ct_3+d}\right)
 \\
 = e^{-\frac{k-2}{2}\left[r_{12}(\sgn(t_1-\frac{d}{c})-\sgn(t_2-\frac{d}{c})) + r_{23}(\sgn(t_2-\frac{d}{c}) -\sgn(t_3-\frac{d}{c})) +r_{31}(\sgn(t_3-\frac{d}{c}) -\sgn(t_1-\frac{d}{c}))\right]}
 \ \Omega_3\ . 
\end{multline}
The general solution is found with the help of the identity (\ref{signs}),
\bea
\boxed{
\Omega_3=|t_{12}|^{\ell_{12}^3}|t_{13}|^{\ell_{13}^2} |t_{23}|^{\ell_{23}^1} e^{\frac{k-2}{2}\left[ r_{12}\sgn t_{12} +r_{23}\sgn t_{23} +r_{31}\sgn t_{31}\right]}\ C_{\sgn t_{12}t_{23}t_{31}} }\ ,
\label{omt}
\eea
where $C_\lambda$ 
%$C_\lambda\left(\ell_1\underset{r_{12}}{|}\ell_2\underset{r_{23}}{|} \ell_3\underset{r_{31}}{|}\right)$ 
is an arbitrary function of the $\USLR$-invariant combination $\lambda=\sgn t_{12}t_{23}t_{31}=\pm$. Thus, the boundary three-point function is written in terms of two independent structure constants $C_\pm$. This reflects the fact that the tensor product of two continuous representation contains two copies of each continuous representation.

Notice that $r_{12},r_{23},r_{31},C_\pm$ cannot be unambiguously determined from $\Omega_3$. The ambiguity corresponds to the invariance of $\Omega_3$ under $r_{ij}\rar r_{ij}+r_0,\ C_\lambda \rar e^{\frac{k-2}{2} r_0 \lambda } C_\lambda$, which follows from the identity (\ref{signs}). This ambiguity will be relevant in the comparison between the exact three-point function and the geometrical prediction. 
 
\subsection{Fourier transformation to the $\nu$-basis}

The first aim of the next section will be to check that the $\Hp$ boundary three-point function predicted by the $\Hp$-Liouville relation is of the form (\ref{omt}) dictated by the $\SLR$ symmetry. However, the $\Hp$-Liouville relation will not directly yield the boundary three-point function $\Omega_3$ of the $t$-basis fields $ _r\Psi^\ell(t|w)_{r'}$ used so far, but rather the following $\nu$-basis boundary three-point function
\bea
\Ome = \prod_{i=1}^3\left( |\nu_i|^{\ell_i+1} \int_\R dt_i\ e^{i\nu_i t_i}\right) \Omega_3 = \la {}_{r_{31}} \Psi^{\ell_1}(\nu_1|w_1)_{r_{12}} \Psi^{\ell_2}(\nu_2|w_2)_{r_{23}} \Psi^{\ell_3}(\nu_3|w_3)_{r_{31}}\ra\ ,
\label{iiio}
\eea
where the $\nu$-basis boundary fields are defined as
\bea
{}_r\Psi^{\ell}(\nu|w)_{r'} = |\nu|^{\ell+1} \int_\R dt\ e^{i\nu t}\  {}_r\Psi^{\ell}(t|w)_{r'} \ , \qquad\nu\in \R\ .
\eea
The present subsection is therefore devoted to the technical task of 
computing $\Ome$ by straightforward Fourier transformation of the $t$-basis result (\ref{omt}), which amounts to formulating the $\USLR$ symmetry constraint in the $\nu$-basis. 

\paragraph{Properties of the $\nu$-basis.}
Only two of the three independent $\USLR$ symmetries have a simple action on $\nu$-basis fields. The first one is $t$-translation symmetry, which implies $\nu$ conservation, so that the $\nu$-basis three-point function $\Ome$ must have a $\delta(\nu_1+\nu_2+\nu_3)$ factor. The second one is $t$-dilatation symmetry, which corresponds to $\nu$-dilatation symmetry, and implies that $\Ome$ is a nontrivial function of only one dilatation-invariant real variable, say $z=-\frac{\nu_1}{\nu_2}\in \R$. Note however that only positive dilatations are allowed, namely $\nu_i\rar \al \nu_i$ with $\al>0$. The nontriviality of the transformation $\nu_i\rar -\nu_i$ implies that $\Ome$ should be thought of as a function on a double cover of $\R$: 
\bea
\begin{array}{|c|ccccccccccccc|}
\hline
z & \infty\! & &\! 0\! & &\! 1\! & &\! \infty\! & &\! 0 \! & &\! 1\! & &\! \infty
\\
\hline
\sgn (\nu_1,\nu_2,\nu_3) & & (--+) & & (+-+) & & (+--) & & (++-) & &
(-+-) & & (-++) & 
\\
\hline
{\rm Notation} & & [+312] & & [-231]  & & [+123] & & [-312] & & [+231] & & [-123] &
\\
\hline
\end{array}
\label{modsp}
\eea
The notation for a regime of $\sgn(\nu_1,\nu_2,\nu_3)$ starts with $\sgn\nu_1\nu_2\nu_3=\pm$, and then indicates the order of the fields on the worldsheet boundary, starting with the index $i$ such that $\sgn\nu_i=\sgn\nu_1\nu_2\nu_3$.

Let me describe more precisely the $\nu$-dependence of $\Ome$. As will follow from the direct calculation of $\Ome$, and could alternatively be derived from the 
local $s\ell(2,\R)$ symmetry, $\Ome$ is a linear combination of hypergeometric functions of the type:
\bea
\begin{array}{|l|} 
\hline
  \begin{array}{r}  {\cal F}^{(3)}_\eta \equiv  \delta(\tsum\nu_i)
 |\nu_1|^{-\ell_1-\ell_3^\eta-1}|\nu_2|^{\ell_2+1}|\nu_3|^{\ell_3^\eta+1}
 F(\ell_{123^\eta}+2,\ell_{23^\eta}^1+1,2\ell_3^\eta+2,-\frac{\nu_3}{\nu_1})
\\  = \delta(\tsum \nu_i)
|\nu_1|^{\ell_1+1} |\nu_2|^{-\ell_2-\ell_3^\eta-1} |\nu_3|^{\ell_3^\eta+1}
F(\ell_{123^\eta}+2,\ell_{13^\eta}^2+1,2\ell_3^\eta+2,-\frac{\nu_3}{\nu_2}) 
 \end{array}
\\
\hline
 \begin{array}{r}{\cal F}^{(2)}_\eta \equiv \delta(\tsum\nu_i)
 |\nu_1|^{\ell_1+1}|\nu_2|^{\ell_2^\eta+1}|\nu_3|^{-\ell_1-\ell_2^\eta-1}
F(\ell_{12^\eta3}+2,\ell_{12^\eta}^3+1,2\ell_2^\eta+2,-\frac{\nu_2}{\nu_3})  
\\ = \delta(\tsum\nu_i)
|\nu_1|^{-\ell_2^\eta-\ell_3-1} |\nu_2|^{\ell_2^\eta+1} |\nu_3|^{\ell_3+1}
F(\ell_{12^\eta 3}+2,\ell_{2^\eta 3}^1 +1,2\ell_2^\eta+2, -\frac{\nu_2}{\nu_1})
\end{array}
\\
\hline
\begin{array}{r} {\cal F}^{(1)}_\eta\equiv \delta(\tsum\nu_i)
|\nu_1|^{\ell_1^\eta+1}|\nu_2|^{-\ell_1^\eta-\ell_3-1} |\nu_3|^{\ell_3+1}
F(\ell_{1^\eta 23}+2,\ell_{1^\eta 3}^2+1,2\ell_1^\eta+2,-\frac{\nu_1}{\nu_2})  \\
=  \delta(\tsum\nu_i) 
|\nu_1|^{\ell_1^\eta+1}|\nu_2|^{\ell_2+1} |\nu_3|^{-\ell_1^\eta-\ell_2-1}
F(\ell_{1^\eta 23}+2,\ell_{1^\eta 2}^3+1,2\ell_1^\eta+2,-\frac{\nu_1}{\nu_3}) \end{array}
\\
\hline
\end{array}
\label{fff}
\eea
where $\eta=\pm$ and $\ell^+=\ell,\ell^-=-\ell-1$ thus $\ell_{12^\eta}^3 = \ell_1+\ell_2^\eta-\ell_3$. The arguments of the hypergeometric functions are assumed to belong to $]-\infty,1[$, which happens for ${\cal F}^{(3)}_\eta$ provided $\nu_1\nu_2<0$. (In particular, ${\cal F}^{(3)}_\eta$ has a power-like behaviour near $\nu_3=0$, but behaves as a linear combination of powers of $|\nu_1|$ and $|\nu_2|$ near $\nu_1=0$ and $\nu_2=0$ respectively.) 
Therefore, out of the three alternative bases ${\cal F}^{(1)}_\eta,{\cal F}^{(2)}_\eta,{\cal F}^{(3)}_\eta$, only two can be used for given values of $\nu_1,\nu_2,\nu_3$. For instance, in the regimes $[\pm 312]$, the two bases ${\cal F}^{(1)}_\eta,{\cal F}^{(2)}_\eta$. 

So the $\nu$-basis three-point function $\Ome$
should have expressions of the form
\bea
\Ome = \sum_{\lambda=\pm} C_\lambda \sum_{\eta=\pm} T^{[\sgn\nu_i]}_{\lambda,\eta} {\cal F}^{(j)}_{\eta}\ ,
\label{ctf}
\eea
where $[\sgn\nu_i]$ denotes a regime, for instance $[+312]$, and $j$ denotes one of the two allowed bases in that regime, here $j=1,2$. Depending on this choice of basis, the coefficient will be denoted as 
$T^{[+3(1)2]}_{\lambda,\eta}$ or $T^{[+31(2)]}_{\lambda,\eta}$. These coefficients relate the
$\nu$-basis three-point structure constants $\Cnu_\eta^{[\sgn\nu_i]} = \sum_{\lambda=\pm} C_\lambda  T^{[\sgn\nu_i]}_{\lambda,\eta}$, which depend on the choices of regime and basis, to the $t$-basis three-point structure constants $C_\lambda$, which do not.  

\paragraph{Calculation of $\Ome$.}

Let me explicitly demonstrate that $\Ome$ indeed has an expression of the form (\ref{ctf}), and determine the coefficients $T_{\lambda,\eta}$, by computing the integral (\ref{iiio}). This integral can be split into six terms corresponding to the six possible orderings of $t_1,t_2,t_3$ on the real line. Up to a global $r_{ij}$-dependent factor, the ordering $t_1<t_2<t_3$ yields the following term:
\bea
J_{123}\equiv \prod_{i=1}^3|\nu_i|^{\ell_i+1}\int_{t_1<t_2<t_3} dt_1\ dt_2\ dt_3\ e^{i(\nu_1t_1+\nu_2t_2+\nu_3t_3)} |t_{12}|^{\ell_{12}^3} |t_{23}|^{\ell_{23}^1} |t_{13}|^{\ell_{13}^2} \ .
\label{jint}
\eea
Introduce a variable $u$ by $|t_{13}|^{\ell_{13}^2} = \frac{1}{\G(-\ell_{13}^2)} \int_0^\infty du\ e^{-u|t_1-t_3|} u^{-\ell_{13}^2-1}$. Shift $t_1\rar t_1+t_2$ and $t_3\rar t_3+t_2$, then integrate over $t_i$, and find
\begin{multline}
J_{123} = \delta(\nu_1+\nu_2+\nu_3)\prod_{i=1}^3|\nu_i|^{\ell_i+1}
 \\ \times
 \frac{\G(\ell_{12}^3+1)\G(\ell_{23}^1+1)}{\G(-\ell_{13}^2)}  \int_0^\infty du\ u^{-\ell_{13}^2-1} (u+i\nu_1)^{-\ell_{12}^3-1} (u-i\nu_3)^{-\ell_{23}^1-1} \ .
\end{multline}
The result is an hypergeometric function \cite{gr65}, which is a priori ambiguous when its (real) argument belongs to $]1,\infty[$. In this case, by construction, the hypergeometric function is
determined by analytic continuation from the region $i\nu_1,-i\nu_3\in \R_+$. This understood, the result can be written as
\bea
J_{123}&=&\G(\ell_{12}^3+1)\G(\ell_{23}^1+1)\frac{\G(\ell_{123}+2)}{\G(2\ell_2+2)} e^{i\frac{\pi}{2}(\ell_{123}+2)\sgn\nu_3} 
{\cal F}^{(2)}_+ \ .
\eea
Now consider all six terms contributing to the integral (\ref{iiio}) in the 
regime $[\sigma 123]$ with $\sigma = \sgn\nu_1=-\sgn\nu_2=-\sgn\nu_3$. The four terms $J_{123},J_{132},J_{231},J_{321}$ yield ``good'' hypergeometric functions ${\cal F}_+^{(2)},{\cal F}_+^{(3)}$ with arguments in $]-\infty,1[$, whereas the two remaining integrals $J_{213},J_{312}$ yield ``bad'' hypergeometric functions ${\cal F}_+^{(1)}$ with arguments in $]1,\infty[$. These can however be unambiguously rewritten as combinations of either ${\cal F}^{(2)}_\pm$ or ${\cal F}^{(3)}_\pm$ functions. The end result is $\Ome = \sum_\lambda C_\lambda \sum_\eta T^{[\sigma 1(2)3]}_{\lambda,\eta} {\cal F}^{(2)}_\eta$, with the blocks ${\cal F}^{(2)}_\eta$ of eq. (\ref{fff}) and the coefficients
% Change of convention \sigma -> -\sigma wrt draft! 
\begin{multline}
T^{[\sigma 1(2)3]}_{\lambda,+} = \frac18\G(\ell_{123}+2)\G(\ell^1_{23}+1)
\G(\ell_{12}^3+1) \G(-2\ell_2-1)
e^{i\lambda\sigma\frac{\pi}{2}\ell_{123}} 
\label{tlp}
\\ \times
\left[ e^{\lambda\frac{r_{23}-r_{31}-r_{12}}{2b^2}} 
\sin \pi \ell^1_{23}
   + e^{\lambda\frac{r_{12}-r_{23}-r_{31}}{2b^2}} \sin\pi \ell^3_{12} 
  - e^{\lambda\frac{r_{31}-r_{12}-r_{23}}{2b^2}}  e^{-i\lambda\sigma \pi\ell_{123}} \sin 2\pi
  \ell_2 \right]\ ,
\end{multline}
\bea
T^{[\sigma 1(2)3]}_{\lambda,-} = -\frac{\pi }{4}\G(\ell^2_{13}+1)\G(2\ell_2+1)
e^{-\lambda \frac{r_{31}}{2b^2}} e^{i\lambda\sigma\frac{\pi}{2}(\ell^2_{13}+1)} 
\sin\left(\pi \ell_2 +i\sigma \frac{r_{23}-r_{12}}{2b^2}\right)\ .
\label{tlm}
\eea
This completes the computation of the Fourier transform $\Ome$ of the general solution $\Omega_3$ (\ref{omt}) of the $\USLR$ symmetry condition. The coefficients $T_{\lambda,\eta}$ which appear in the result will play an important role in the following, so let me study some of their properties. 

\paragraph{Some properties of the coefficients $T_{\lambda,\eta}$.} 
The determinant of the $2\times 2$ matrix $T_{\lambda,\eta}$ is
\begin{multline}
\det T^{[\sigma 1(2)3]} = \frac{i\pi^2\sigma}{8(2\ell_2+1)} \G(\ell_{12}^3+1) \G(\ell_{13}^2+1) \G(\ell_{23}^1+1) \G(\ell_{123}+2) 
\\ \times \sin\left(\pi \ell_1 +i\sigma\frac{r_{31}-r_{12}}{2b^2}\right) \sin\left(\pi \ell_2 +i\sigma \frac{r_{23}-r_{12}}{2b^2}\right) \sin \left(\pi \ell_3 +i\sigma\frac{r_{31}-r_{23}}{2b^2}\right)\ ,
\end{multline}
and its inverse $\left(T^{-1}\right)_{\lambda,\eta} = \frac{\eta\lambda}{\det T} T_{-\eta,-\lambda}$. 

The existence of the two bases ${\cal F}^{(2)}_\pm$ and ${\cal F}^{(3)}_\pm$ means $\sum_\eta T^{[\sigma 1(2)3]}_{\lambda,\eta} {\cal F}^{(2)}_\eta = \sum_{\eta'} T^{[\sigma 12(3)]}_{\lambda,\eta'} {\cal F}^{(3)}_{\eta'}$. 
Given the relations $\sum_{\eta} {\cal F}^{(i)}_\eta M^{(ij)k}_{\eta\eta'} = {\cal F}^{(j)}_{\eta'}$ between the two bases of conformal blocks ${\cal F}^{(i)},{\cal F}^{(j)}$ in regimes where $\sgn\nu_i=\sgn\nu_j$, this implies relations of the type 
\bea
T^{[\sigma 1(2)3]}_{\lambda,\eta} = \sum_{\eta'} M_{\eta\eta'}^{(23)1} T^{[\sigma 12(3)]}_{\lambda,\eta'}
 \ , \label{tmt}
 \eea
 where the monodromy matrix is
\bea
M_{\eta\eta'}^{(23)1} = 
\frac{\G(2\ell_3^{\eta'}+2)
  \G(-2\ell_2^\eta-1)}{ \G(1+\ell_1-\ell_2^\eta+\ell_3^{\eta'})
  \G(-\ell_1-\ell_2^\eta+\ell_3^{\eta'})} \ ,\quad \eta,\eta'=\pm\ . 
  \label{mee}
  \eea
(Such relations can be explicitly checked using $T^{[\sigma 12(3)]}_{\lambda,\eta}= T^{[-\sigma 1(3)2]}_{\lambda,\eta}$.)

\paragraph{$\USLR$ symmetry condition in the $\nu$-basis.}

Finally, examining the coefficients $T_{\lambda,\eta}$ yields the $\nu$-basis formulation of the $\USLR$ symmetry condition, that is the formulation which will be used in the next section. 
The global structure of the symmetry group $\USLR$ is actually encoded in the behaviour
of $\Ome$ when each of the $\nu_i$ vanish, say $\nu_2=0$. Such a point separates two regimes where the ${\cal F}^{(2)}_\eta$ basis can be used, say $[\sigma 123]$ and $[-\sigma 312]$. It turns out that the coefficient $T_{\lambda,+}$ is continuous across this singularity, whereas $T_{\lambda,-}$ has a jump:
\bea
T_{\lambda,+}^{[\sigma 1(2)3]} = T_{\lambda,+}^{[-\sigma 31(2)]} \scs T_{\lambda,-}^{[\sigma 1(2)3]} = \frac{\sin\left(\pi \ell_2 +i\sigma \frac{r_{23}-r_{12}}{2b^2}\right)}{\sin\left(\pi \ell_2 -i\sigma \frac{r_{23}-r_{12}}{2b^2}\right)}  T_{\lambda,-}^{[-\sigma 31(2)]}\ .
\label{tptm}
\eea
Since this does not depend on $\lambda$, this can be interpreted as the jump condition on the $\nu$-basis three-point structure constants $\Cnu_\eta^{[\sgn\nu_i]} = \sum_{\lambda=\pm} C_\lambda  T^{[\sgn\nu_i]}_{\lambda,\eta}$.
Thus, $\USLR$ symmetry relates the $\nu$-basis structure constants in the six regimes (\ref{modsp}). Only two of these structure constants are independent, as is expected from their relation with the two $t$-basis structure constants $C_\lambda$.  

\section{The three-point function: explicit calculation}

The symmetry properties of the three-point function, in other words the kinematics, leave the two structure constants $C_\lambda$ in (\ref{omt}) undetermined. The geometrical calculation only gives very partial information on these structure constants. A full determination requires a more powerful dynamical principle. The principle which I will now use is the relation of the $\Hp$ model with Liouville theory \cite{rt05,hr06}. The boundary three-point function following from this principle leads to a crossing-symmetric four-point function \cite{hr06}. The agreement of the $\Hp$-Liouville relation with the $\USLR$ symmetry analysis and with the geometrical calculation is however not obvious, and will have to be checked explicitly.

\subsection{The three-point function from Liouville theory}
 
The $\Hp$-Liouville relation predicts all correlators of the $\Hp$ model on a disc in terms of correlators of Liouville theory on a disc. In this subsection I will review this prediction in the particular case of the $\Hp$ boundary three-point function, and show that in this case the relevant Liouville correlators can be explicitly computed.

\paragraph{Prediction of the boundary three-point function.} 
According to \cite{hr06}, 
\bea
\Ome = \delta(\tsum\nu_i) |\tsum \nu_iw_i|^{1+\frac{3}{2b^3}} |\nu_1\nu_2\nu_3w_{12}w_{23}w_{31}|^{-\frac{1}{2b^2}} \la B^{\beta_1}(w_1) B^{\beta_2}(w_2) B^{\beta_3}(w_3) B^{-\frac{1}{2b}}(y) \ra\ .
\label{obbbb}
\eea
The correlator is a disc boundary four-point function in Liouville theory at central charge $c_L=1+6Q^2$ with $Q=b+b^{-1}$ and $b^2=\frac{1}{k-2}$, which involves three boundary fields of momenta $\beta_i=b(\ell_i+1)+\frac{1}{2b}$ and conformal weight $\beta_i(Q-\beta_i)$, together with one  degenerate boundary  field of momentum $-\frac{1}{2b}$, whose position $y=-\frac{\nu_1w_2w_3+\nu_2w_3w_1+\nu_3w_1w_2}{\nu_1w_1+\nu_2w_2+\nu_3w_3} $ is more elegantly defined as
\bea
\varphi(y)=0\qquad {\rm where}\qquad \varphi(y)\equiv\sum_i\frac{\nu_i}{y-w_i} \ .
\label{vpy}
\eea
The degenerate field $B^{-\frac{1}{2b}}(y)$ needs not always be located between $w_3$ and $w_1$ as in (\ref{obbbb}), but can live at any position on the worldsheed boundary, depending on the variables $\nu_i$: more precisely, between fields at $w_i$ and $w_j$ if and only if $\nu_i\nu_j>0$. 
The behaviour of Liouville theory on the boundary of the worldsheet is assumed to be characterized by so-called FZZT branes \cite{fzz00,tes00}. The parameter of the FZZT brane at a point $w$ of the boundary is assumed to be
 \footnote{The convention for the Liouville boundary parameter $s$ is that the boundary cosmological constant is proportional to $\cosh 2\pi b s$.} 
\bea
s=\frac{r}{2\pi b} - \frac{i}{4b}\sgn \varphi(w)\ ,
\label{srphi}
\eea
where $r$ is the $\Hp$ model's boundary parameter ($r_{12}$,$r_{23}$ or $r_{31}$) at the same point $w$. In the regime $[+123]$ i.e. $\nu_2,\nu_3<0<\nu_1$ the worldsheet looks like
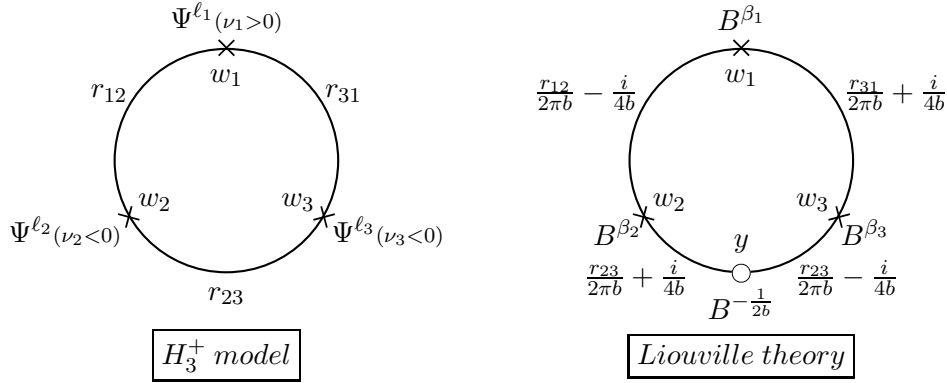
\begin{center}
\begin{tabular}{ccc}
 \psset{unit=.5cm}
 \pspicture[](-4,-4.5)(4,4)
 \pscircle(0,0){3}
 \psdots[dotstyle=+,dotangle=45,dotscale=2](0,3)
 \rput{*0}(0,3.8){$\Psi^{\ell_1}\scriptstyle(\nu_1>0)$} \rput{*0}(0,2.2){$w_1$}
 \rput{-120}{
 \psdots[dotstyle=+,dotangle=45,dotscale=2](0,3)
 \rput{*0}(0,3.8){\rput(1,0){$\Psi^{\ell_3}\scriptstyle (\nu_3<0)$}} \rput{*0}(0,2.2){$w_3$} 
 }
 \rput{120}{
 \psdots[dotstyle=+,dotangle=45,dotscale=2](0,3)
 \rput{*0}(0,3.8){\rput(-1,0){$\Psi^{\ell_2}\scriptstyle(\nu_2<0)$}} \rput{*0}(0,2.2){$w_2$} 
 }
 \rput{-60}{\rput{*0}(0,3.6){$r_{31}$}}
 \rput{60}{\rput{*0}(0,3.6){$r_{12}$}}
 \rput(0,-3.6){$r_{23}$}
 \endpspicture
 & \hspace{2cm} &
 \psset{unit=.5cm}
 \pspicture[](-4,-4.5)(4,4)
 \pscircle(0,0){3}
 \psdots[dotstyle=+,dotangle=45,dotscale=2](0,3)
 \rput{*0}(0,3.8){$B^{\beta_1}$} \rput{*0}(0,2.2){$w_1$}
 \rput{-120}{
 \psdots[dotstyle=+,dotangle=45,dotscale=2](0,3)
 \rput{*0}(0,3.8){$B^{\beta_3}$} \rput{*0}(0,2.2){$w_3$} 
 }
 \rput{120}{
 \psdots[dotstyle=+,dotangle=45,dotscale=2](0,3)
 \rput{*0}(0,3.8){$B^{\beta_2}$} \rput{*0}(0,2.2){$w_2$} 
 }
 \rput{-60}{\rput{*0}(0,3.6){\rput(1,0){$\frac{r_{31}}{2\pi b}+\frac{i}{4b}$}}}
 \rput{60}{\rput{*0}(0,3.6){\rput(-1,0){$\frac{r_{12}}{2\pi b}-\frac{i}{4b}$}}}
 \rput{180}{\psdots[dotstyle=o,dotangle=45,dotscale=2](0,3)
 \rput{*0}(0,3.8){$B^{-\frac{1}{2b}}$} \rput{*0}(0,2.2){$y$} }
 \rput{-150}{\rput{*0}(0,3.6){\rput(1,0){$\frac{r_{23}}{2\pi b}-\frac{i}{4b}$}}}
 \rput{150}{\rput{*0}(0,3.6){\rput(-1,0){$\frac{r_{23}}{2\pi b}+\frac{i}{4b}$}}}
 \endpspicture
  \\ \boxed{\Hp\ model} & & \boxed{Liouville\ theory} 
\end{tabular}
\end{center}

\paragraph{Calculation of the relevant Liouville four-point function.}

Due to the presence of the degenerate field $B^{-\frac{1}{2b}}$, the four-point function in eq. (\ref{obbbb}) obeys a second-order differential equation \cite{bpz84}. The conformal blocks which solve this equation are, up to power factors, hypergeometric functions of cross-ratios of the type $\frac{(y-w_2)(w_1-w_3)}{(y-w_3)(w_1-w_2)} = -\frac{\nu_2}{\nu_3}$. It can be checked
\footnote{A similar calculation was written explicitly in \cite{rt05} in the case of the relation between the $\Hp$ three-point function and the Liouville four-point function {\it on a sphere}.}
that these hypergeometric solutions, combined with the extra factors in eq. (\ref{obbbb}), yield the functions ${\cal F}^{(i)}_\eta$ (\ref{fff}). (Here and in the following I omit the $w$-dependence of the $\Hp$ three-point function.) The two alternative bases of conformal blocks for given values of $\sgn\nu_i$ correspond to two possible decompositions of the Liouville four-point function. For instance, if $w_2<y<w_3$ then the field $B^{-\frac{1}{2b}}(y)$ can be associated with either $B^{\beta_2}(w_2)$ or $B^{\beta_3}(w_3)$. In the former case, this means choosing the basis of conformal blocks ${\cal F}^{(2)}_\eta$, such that each block ${\cal F}^{(2)}_\pm$ has a power-like behaviour in the limit $y\rar w_2 \Leftrightarrow \nu_2\rar 0$. This basis has two elements $\eta=\pm$, which correspond to the two fusion channels $B^{-\frac{1}{2b}} \times B^{\beta_2} \rar \sum_{\eta=\pm} B^{\beta_2-\frac{\eta}{2b}}$. The corresponding Liouville conformal blocks can be drawn as follows:
\bea
{\cal F}^{(2)}_\eta\propto
\psset{unit=.3cm}
\pspicture[](-4,-3.5)(4,3.2)
\psline(2,3)(0,2)(0,-2)(-2,-3)
\thecoil(2,-3)(0,-2)\psline(-2,3)(0,2)
\rput(2.9,3.2){$\beta_3$} \rput(3.1,-3.2){$-\tfrac{1}{2b}$} \rput(-2.9,-3.2){$\beta_2$} \rput(-2.9,3.2){$\beta_1$}
\rput[l](.2,0){$\beta_2-\tfrac{\eta}{2b}$}
\endpspicture
\quad , \hspace{15mm}
{\cal F}^{(3)}_\eta\propto
\psset{unit=.38cm}
\pspicture[](-4,-1.8)(4,1.8)
\psline(3,2)(2,0)(-2,0)(-3,2)
\psline(-3,-2)(-2,0) \thecoil(3,-2)(2,0)
\rput(3.9,2,2){$\beta_3$} \rput(-3.9,2.2){$\beta_1$} \rput(-3.9,-2.2){$\beta_2$} 
\rput(3.9,-2.2){$-\tfrac{1}{2b}$}
\rput(0,.7){$\beta_3-\tfrac{\eta}{2b}$}
\endpspicture
\quad .
\label{fldeg}
\eea

The coefficients of the decomposition of the Liouville four-point function in conformal blocks are certain Liouville structure constants. In the regime $[\sigma 1(2)3]$ with the choice of basis ${\cal F}^{(2)}_\eta$, the $\Hp$ three-point function (\ref{obbbb}) then reads
\begin{multline}
\Ome = \sum_{\eta=\pm}  C^L\left(\beta_1\underset{\frac{r_{12}}{2\pi b} -\sigma\frac{i}{4b}}{|} \beta_2-\tfrac{\eta}{2b} \underset{\frac{r_{23}}{2\pi b}-\sigma\frac{i}{4b}}{|} \beta_3 \underset{\frac{r_{31}}{2\pi b} +\sigma\frac{i}{4b}}{|}\right) 
\\ \times
C^L\left(\beta_2\underset{\frac{r_{23}}{2\pi b}+\sigma\frac{i}{4b}}{|} -\tfrac{1}{2b} \underset{\frac{r_{23}}{2\pi b}-\sigma\frac{i}{4b}}{|} Q-\beta_2+\tfrac{\eta}{2b} \underset{\frac{r_{12}}{2\pi b} -\sigma\frac{i}{4b}}{|} \right) {\cal F}^{(2)}_\eta \ ,
\label{omexpl}
\end{multline}
where the $C^L$ are Liouville three-point structures constants. 

\paragraph{Liouville theory structure constants.}

The Liouville three-point structure constant is explicitly known \cite{pt01} as a function of the 
three momenta $\beta_i$ and the three boundary parameters $s_{ij}$: 
\begin{multline}
C^L\left(\beta_1\underset{s_{12}}{|}\beta_2 \underset{s_{23}}{|} \beta_3 \underset{s_{31}}{|}\right)  = 
 \mu_L^{\frac{Q-\beta_{123}}{2b}} 
\frac{\Gamma_b(2Q-\beta_{123})\Gamma_b(\beta_{23}^1)
 \Gamma_b(Q-\beta_{13}^2)\Gamma_b(Q-\beta_{12}^3)}
{\Gamma_b(Q-2\beta_3)\Gamma_b(Q-2\beta_2)\Gamma_b(Q-2\beta_1)\Gamma_b(Q)}
\\
\times\frac{S_b(Q-\beta_3+is_{31}-is_{23})S_b(Q-\beta_3-is_{23}-is_{31})} 
{S_b(\beta_2+is_{12}-is_{23})S_b(\beta_2-is_{23}-is_{12})} 
\times \frac{1}{i}\int\limits_{Q-i\infty}^{Q+i\infty}dp \;\;
\prod_{i=1}^4 \frac{S_b(U_i+p)}{S_b(V_i+p)} \ ,
\label{clbbb}
\end{multline}
where the special functions $\G_b$ and $S_b$ are described in the Appendix,
$\mu_L$ is the renormalized Liouville cosmological constant, and
the coefficients $U_i$, $V_i$ read
\bea
\begin{array}{lcl}
 U_1 =is_{31}        -\beta_1        &  \hspace{1cm}  & V_1 = -is_{23}-\beta_1+\beta_3 \\
 U_2 = -is_{31}            -\beta_1   &   & V_2 = Q-is_{23}-\beta_1-\beta_3 \\
 U_3 = -Q+\beta_2-          is_{23}       &    &  V_3 = is_{12} \\
 U_4 = -\beta_2-is_{23}               &    &  V_4 = -is_{12}
\end{array}
\eea
In this formula the symmetries of $C^L$ are not manifest: neither
the invariances under permuations of the indices and under individual reflections of boundary parameters $s_{ij}\rar -s_{ij}$, nor the reflection symmetry $C^L\left(\beta_1\underset{s_{12}}{|}\beta_2 \underset{s_{23}}{|} \beta_3 \underset{s_{31}}{|}\right) = R^L_{s_{31},s_{12}}(\beta_1)\ C^L\left(Q-\beta_1\underset{s_{12}}{|}\beta_2 \underset{s_{23}}{|} \beta_3 \underset{s_{31}}{|}\right)$ (where $R^L$ is given in eq. (\ref{rlss})).

The degenerate structure constant $C^L(\beta_2|-\frac{1}{2b}|Q-\beta_2+\frac{\eta}{2b})$ in (\ref{omexpl}) follows from the known formulas \cite{fzz00}
\bea
C^L\left(\beta\underset{s}{|} -\tfrac{1}{2b} \underset{s-\sigma \frac{i}{2b}}{|} Q-\beta+\tfrac{1}{2b} \underset{s'}{|} \right) &=& 1\ ,
\label{clp}
\\
C^L\left(\beta\underset{s}{|} -\tfrac{1}{2b} \underset{s-\sigma \frac{i}{2b}}{|} Q-\beta-\tfrac{1}{2b} \underset{s'}{|} \right) &=& R^L_{s' , s  }(\beta) R^L_{s-\sigma \frac{i}{2b},s'}(Q-\beta-\tfrac{1}{2b})     \ .
\label{clm}
\eea
The first formula is actually a normalization convention, from which the second one is deduced by using the boundary reflection relation ${}_sB^\beta_{s'} = R^L_{s,s'}(\beta)\ {}_s B^{Q-\beta}_{s'} $, where the boundary reflection coefficient is
\bea
R^L_{s,s'}(\beta) = \mu_L^{\frac{Q-2\beta}{2b}} \frac{\G_b(2\beta-Q)}{\G_b(Q-2\beta)} \prod_{\pm,\pm} S_b(Q-\beta\pm is \pm is') \ . 
\label{rlss}
\eea

\subsection{Check of the symmetry \label{subcots}}

The formula (\ref{omexpl}) for the $\nu$-basis three-point function $\Ome$ is explicit but
not particularly illuminating, and it depends on the choices of a particular regime of values of $\nu_i$ and of a particular basis of conformal blocks.
I will now recast it as a formula for 
the $t$-basis structure constants $C_\lambda$ defined in (\ref{omt}), which have no such restrictions and enjoy nicer symmetry properties. 

Before doing this, it is however necessary to show that the explicit formula for $\Ome$ is indeed compatible with the $\USLR$ symmetry which underlies the very definition of $C_\lambda$.
Recall that the $\USLR$ symmetry condition for the boundary three-point function can be formulated as a condition on its behaviour across a singularity of the type $\nu_2=0$, see eq. (\ref{tptm}). So how does the explicit expression (\ref{omexpl}) behave near $\nu_2=0$?

\paragraph{The three-point function $\Ome$ near $\nu_2=0$.}

This amounts to studying the behaviour of the Liouville four-point function in (\ref{obbbb}) near $y=w_2$, at which point the degenerate field $B^{-\frac{1}{2b}}(y)$ crosses the field $B^{\beta_2}(w_2)$. Assuming $\nu_1>0$ and $\nu_3<0$, the worldsheet near $w_2$ then looks like:
\begin{center}
 \begin{tabular}{ccc} 
 \boxed{\nu_2>0}   &  \hspace{1cm}  & \boxed{\nu_2<0} 
 \\
 \psset{unit=.8cm}
 \pspicture[](-4,-1)(4,1)
 \psline(-4,0)(4,0)
 \psdots[dotstyle=+,dotangle=45,dotscale=2](0,0)
 \psdots[dotstyle=o,dotangle=45,dotscale=2](-2,0)
 \rput(-3,-.5){$\frac{r_{12}}{2\pi b}-\frac{i}{4b}$}
 \rput(-1,-.5){$\frac{r_{12}}{2\pi b}+\frac{i}{4b}$}
 \rput(3,-.5){$\frac{r_{23}}{2\pi b}-\frac{i}{4b}$}
 \rput(-2,.5){$B^{-\frac{1}{2b}}$}
 \rput(0,.5){$B^{\beta_2}$}
 \endpspicture
 &
 &
 \psset{unit=.8cm}
 \pspicture[](-4,-1)(4,1)
 \psline(-4,0)(4,0)
 \psdots[dotstyle=+,dotangle=45,dotscale=2](0,0)
 \psdots[dotstyle=o,dotangle=45,dotscale=2](2,0)
 \rput(-3,-.5){$\frac{r_{12}}{2\pi b}-\frac{i}{4b}$}
 \rput(1,-.5){$\frac{r_{23}}{2\pi b}+\frac{i}{4b}$}
 \rput(3,-.5){$\frac{r_{23}}{2\pi b}-\frac{i}{4b}$}
 \rput(2,.5){$B^{-\frac{1}{2b}}$}
 \rput(0,.5){$B^{\beta_2}$}
 \endpspicture
 \end{tabular}
\end{center}
The most complicated factor in (\ref{omexpl}), namely $C^L(\beta_1|\beta_2-\tfrac{\eta}{2b}|\beta_3)$, is actually continuous across $\nu_2=0$. This factor is indeed a Liouville three-point structure constant involving the field $B^{\beta_2-\frac{\eta}{2b}}$ which results from the fusion of $B^{-\frac{1}{2b}}$ and $B^{\beta_2}$: once they have fused, it does not matter which directions the fields came from. On the other hand, the relative positions of the two fields influence the other factor $C^L(\beta_2|-\frac{1}{2b}|Q-\beta_2+\frac{\eta}{2b})$ in the case $\eta=-$, because this factor is then sensitive to the boundary parameter between the two fields, as is clear from eq. (\ref{clm}):
\begin{multline}
\frac{
C^L\left(\beta_2\underset{\frac{r_{23}}{2\pi b}+\sigma\frac{i}{4b}}{|} -\tfrac{1}{2b} \underset{\frac{r_{23}}{2\pi b}-\sigma\frac{i}{4b}}{|} Q-\beta_2+\tfrac{1}{2b} \underset{\frac{r_{12}}{2\pi b} -\sigma\frac{i}{4b}}{|} \right)
}{
C^L\left(\beta_2\underset{\frac{r_{23}}{2\pi b}-\sigma\frac{i}{4b}}{|} 
Q-\beta_2+\tfrac{1}{2b}  \underset{\frac{r_{12}}{2\pi b} -\sigma\frac{i}{4b}}{|}   -\tfrac{1}{2b}\underset{\frac{r_{12}}{2\pi b} +\sigma\frac{i}{4b}}{|} \right)
} \\
= 
\frac{R^L_{\frac{r_{12}}{2\pi b} -\sigma\frac{i}{4b}, \frac{r_{23}}{2\pi b}+\sigma\frac{i}{4b}}(\beta_2)} {R^L_{\frac{r_{12}}{2\pi b} +\sigma\frac{i}{4b}, \frac{r_{23}}{2\pi b}-\sigma\frac{i}{4b}}(\beta_2)} = \frac{\sin\left(\pi \ell_2 +i\sigma \frac{r_{23}-r_{12}}{2b^2}\right)}{\sin\left(\pi \ell_2 -i\sigma \frac{r_{23}-r_{12}}{2b^2}\right)} \ .
\end{multline}
The agreement of this formula with the $\USLR$ symmetry condition eq. (\ref{tptm}) demonstrates the 
consistency of the $\Hp$-Liouville relation for the boundary three-point function with the $\USLR$ symmetry.

\paragraph{Determination of the structure constants $C_\lambda$.}

Let me compare the expression (\ref{omexpl}) of $\Ome$
with the expression (\ref{ctf}) of an $\USLR$-symmetric three-point function in the $\nu$-basis. Many apparently different expressions for $C_\lambda$ can be obtained in the different regimes of $\nu_i$, but they are all guaranteed to be equivalent by the $\USLR$ symmetry. The two regimes $[\pm 123]$ alone yield four equations for the two unknowns $C_\pm$, schematically
\bea
C^L_\sigma(\beta_1|\beta_2-\tfrac{\eta}{2b}|\beta_3) C^L_\sigma(\beta_2|-\tfrac{1}{2b}|Q-\beta_2+\tfrac{\eta}{2b})
= \sum_{\lambda=\pm} C_\lambda T_{\lambda,\eta}^{[\sigma 1(2)3]}, \quad \forall \sigma=\pm,\eta=\pm\ .
\label{ccct}
\eea
A relatively simple formula for $C_\lambda$ is obtained by solving the two equations $(\sigma=\pm,\eta=-)$:
\bea
\boxed{
\begin{array}{l}
 C_\lambda\left(\ell_1\underset{r_{12}}{|}\ell_2 \underset{r_{23}}{|} \ell_3\underset{r_{31}}{|}\right) = -\frac{2}{\pi^3}\G(-\ell^2_{13}) R_{r_{12},r_{23}}(\ell_2)
 \\ 
\hspace{1.2cm} \times \sum_{\sigma=\pm} e^{\lambda\left(\frac{r_{31}}{2b^2}- i\sigma\frac{\pi}{2}\ell^2_{13}\right)}
  C^L\left(\beta_1\underset{\frac{r_{12}}{2\pi b} -\sigma\frac{i}{4b}}{|} Q-\beta_2-\tfrac{1}{2b} \underset{\frac{r_{23}}{2\pi b}-\sigma\frac{i}{4b}}{|} \beta_3 \underset{\frac{r_{31}}{2\pi b} +\sigma\frac{i}{4b}}{|}\right)
\end{array} 
} \ ,
\label{clcl}
\eea
where the $\Hp$ boundary reflection coefficient $R_{r_{12},r_{23}}(\ell_2)$ will shortly be introduced in (\ref{rrr}), the Liouville boundary three-point function $C^L$ is still given by (\ref{clbbb}), 
with Liouville momenta still given by $\beta_i=b(\ell_i+1)+\frac{1}{2b}$. (For a fully explicit formula, see eq. (\ref{fef}) below.)

The manifest symmetry of (\ref{clcl}) under $1\lrar 3$ shows that $C_\lambda$ is invariant not only under cyclic permutations, but under all permutations. Equivalently, the full boundary three-point function $\Omega_3$ (\ref{omt}) is invariant under permutations, combined with $t\rar -t$ in the case of odd permutations. This invariance of $\Omega_3$ follows from the invariance of the Liouville four-point function (\ref{obbbb}) under cyclic permutations and worldsheet parity.

\paragraph{Reflection properties of the three-point function.} 

For the sake of completeness, and also in order to introduce the useful quantities $R_{r,r'}(\ell)$ and $N^\sigma_{r,r'}(\ell)$, let me discuss the reflection of boundary fields and correlators in $\Hp$.
By reflection I mean the relation between fields of spins $\ell$ and $-\ell-1$, which transform in the same representation of $\USLR$. The reflection of the $t$-basis boundary field
\footnote{
Knowing the reflection behaviour of fields is equivalent to knowing the boundary two-point function \cite{hr06} 
\bea
\la {}_r\Psi^{\ell_1}(t_1|w_1)_{r'} \Psi^{\ell_2}(t_2|w_2)_r \ra 
= \delta(\ell_1+\ell_2+1)
\delta(t_{12}) + \delta(\ell_1-\ell_2) \tilde{R}_{r,r'}^{H}(\ell_1)
|t_{12}|^{2\ell_1} e^{\frac12(k-2)(r-r')\sgn t_{12}} \ .
\label{twopt}
\eea
}
is fairly complicated in that it involves an integral over the isospin variable $t$,
\bea
{}_r \Psi^\ell(t|w)_{r'} = R_{r,r'}(\ell) \int_\R dt'\ |t-t'|^{2\ell} e^{-\frac{k-2}{2}(r-r')\sgn(t-t')}\ {}_r\Psi ^{-\ell-1}(t'|w)_{r'} \ ,
\eea
with the $t$-basis reflection number (which is invariant under $r\lrar r'$) 
% (correcting a typo in \cite{hr06})
\bea
R_{r,r'}(\ell) = 
N_{r,r'}^\sigma(\ell)
R^L_{ \frac{r}{2\pi b} +\sigma\frac{i}{4b} , \frac{r'}{2\pi b}-\sigma\frac{i}{4b}}
\left(\beta\right) 
\ \ {\rm with}\ \ N_{r,r'}^\sigma(\ell) = \frac{\pi}{\G(2\ell+1)} \frac{1}{\sin(\pi \ell+i\sigma\frac{r-r'}{2b^2})} \ ,
\label{rrr}
\eea
where $\beta=b(\ell+1)+\frac{1}{2b}$. 
The behaviour of $C_\lambda$ under reflection can in principle be directly deduced from the behaviour of individual boundary fields. It is however simpler to formulate the problem in the $\nu$-basis, which (as follows from the $\Hp$-Liouville relation) actually diagonalizes reflection:
\bea
{}_r \Psi^\ell(\nu|w)_{r'} = 
R^L_{\frac{r}{2\pi b}+\frac{i}{4b}\sgn\nu,\frac{r'}{2\pi b}-\frac{i}{4b}\sgn \nu}\left(\beta \right)
%\frac{1}{\pi}\G(2\ell+1)\sin\left(\pi \ell+i \frac{r-r'}{2b^2}\sgn \nu\right) R_{r,r'}(\ell)
\ \ {}_r \Psi^{-\ell-1}(\nu|w)_{r'} \ .
\label{refnu}
\eea
A third way to deduce the reflection of $C_\lambda$ is to directly use their expression in terms of the (reflection-friendly) Liouville structure constants (\ref{ccct}). The result is
\bea
C_\lambda \left(\ell_1\underset{r_{12}}{|}\ell_2 \underset{r_{23}}{|} \ell_3\underset{r_{31}}{|}\right)
= \sum_{\lambda'} R_{\lambda\lambda'}^{(2)} \left(\ell_1\underset{r_{12}}{|}\ell_2 \underset{r_{23}}{|} \ell_3\right) C_{\lambda'} \left(\ell_1\underset{r_{12}}{|}-\ell_2-1 \underset{r_{23}}{|} \ell_3\underset{r_{31}}{|}\right) \ ,
\eea
where the ($r_{31}$-independent) reflection matrix for the spin $\ell_2$ is 
\begin{multline}
R^{(2)} = -\frac{1}{2\pi^2} \G(2\ell_2+1) \G(-\ell_{12}^3)\G(-\ell_{23}^1) R_{r_{12},r_{23}}(\ell_2)
\\ 
\times \left(\begin{array}{ccc} e^{-\frac{r_{12}-r_{23}}{2b^2}} \sin \pi \ell_{12}^3 +e^{\frac{r_{12}-r_{23}}{2b^2}} \sin \pi \ell_{23}^1 & &e^\frac{r_{12}+r_{23}}{2b^2} \sin 2\pi \ell_2 
              \\ e^{-\frac{r_{12}+r_{23}}{2b^2}}\sin 2\pi \ell_2 & \ \ & e^{\frac{r_{12}-r_{23}}{2b^2}} \sin \pi \ell_{12}^3 +e^{-\frac{r_{12}-r_{23}}{2b^2}} \sin \pi \ell_{23}^1
             \end{array}\right)\ .
\end{multline}

\subsection{Check of the geometrical limit\label{subgeom}}

Let me now compute the geometrical limit of the $\Hp$ three-point function in order to compare it with the prediction of subsection \ref{submini}. This amounts to taking the level $k$ to infinity (equivalently $b=(k-2)^{-\frac12}\rar 0$), while keeping the spins $\ell_i$ fixed, and the boundary parameters $r_{ij}$ fixed and equal to a common value $r$. Let me perform this limit on the explicit expression for the
boundary three-point structure constant (\ref{clcl}), 
\begin{multline}
 C_\lambda = \frac{4}{i\pi ^2} \left(\frac{\mu_L}{b^2}\right)^{-\frac{\ell_{123}+2}{2}}  \frac{\G_b(-b\ell_{13}^2)\G_b(-b\ell_{12}^3) \G_b(-b(\ell_{123}+2)) \G_b(Q+b\ell_{23}^1)}{ \G_b(Q) \prod_{i=1}^3 \G_b(-b(2\ell_i+1))}  
 \\ \times 
 \sum_{\sigma=\pm}  e^{\lambda\left(\frac{r_{31}}{2b^2}+ i\sigma\frac{\pi}{2}\ell^2_{13}\right)} 
 \frac{S_b(\tfrac{1}{2b}+i\sigma\tfrac{r_{12}+r_{23}}{2\pi b} -b\ell_2) S_b(i\sigma\tfrac{r_{23}-r_{12}}{2\pi b} -b\ell_2) }{ 
 S_b(\tfrac{1}{2b}+i\sigma\tfrac{r_{23}+r_{31}}{2\pi b}+b(\ell_3+1)) S_b(i\sigma\tfrac{r_{23}-r_{31}}{2\pi b} +b(\ell_3+1))}  \int dp\ 
\\
\frac{S_b(\tfrac{1}{2b} +i\sigma\tfrac{r_{23}+r_{31}}{2\pi b}-b\ell_1 +bp) S_b(i\sigma\tfrac{r_{23}-r_{31}}{2\pi b} -b\ell_1+bp) S_b(-b\ell_2+bp) S_b(Q+b\ell_2+bp)} 
{\prod_\pm S_b(Q+b(\ell_3^\pm -\ell_1)+bp) S_b(\tfrac{1}{2b}+i\sigma\tfrac{r_{12}+r_{23}}{2\pi b}+b +bp) S_b(Q+i\sigma\tfrac{r_{23}-r_{12}}{2\pi b} +bp)} \ .
\label{fef}
\end{multline}

\paragraph{Limits of $C_\lambda$ and $\Omega_3$.}

The behaviour of the special function $S_b$ as $b\rar 0$ is given in eqs. (\ref{limsb},\ref{limsbr}). 
The argument of the function $S_b$ must behave in certain ways for the limit to exist. In the geometrical limit, the spins $\ell_i$ and brane parameters $r_{ij}$ are kept fixed. This allows $C_\lambda$ to have a well-defined limit only provided all brane parameters are equal, as was anticipated on more heuristic grounds in subsection \ref{submini}. Calling $r$ this common parameter, and neglecting some numerical prefactors, the limit is found by direct calculation to be
\bea
C_\lambda\left(\ell_1\underset{r}{|}\ell_2 \underset{r}{|}\ell_3\underset{r}{|}\right) \underset{b\rar 0}{\sim} (\cosh r)^{\sum\ell_i+2}\ e^{\lambda \frac{r}{2b^2}}\ C^0 \ ,
\eea
where the constant $C^0$, which depends only on the spins $\ell_1,\ell_2,\ell_3$, is
\bea
C^0 &=&  \frac{\G(-\ell_{13}^2) \G(-\ell_{12}^3) \G(-\ell_{123}-1)}{\prod_{i=1}^3 \G(-2\ell_i-1)}\ \frac{\G(-\ell_2)}{\G(\ell_3+1)}\ \cos\tfrac{\pi}{2}\ell_{13}^2 \times I\ ,
\\ 
I &\equiv & \int dp\ \frac{\G(-p)\G(\ell_1-\ell_2-p) \G(-\ell_1+p) \G(-\ell_3+p)\G(\ell_3+1+p)}{\G(-\ell_1-\ell_2+p)}\ .
\label{iint}
\eea
Now insert this into the three-point function $\Omega_3$, eq. (\ref{omt}), and use formula (\ref{signs}) to get the simple result
\bea
\Omega_3&\underset{b\rar 0}{\sim}& 
|t_{12}|^{\ell_{12}^3}|t_{13}|^{\ell_{13}^2} |t_{23}|^{\ell_{23}^1} (\cosh r)^{\sum\ell_i+2} C^0\ .
\label{otco}
\eea
The dependences on $r$ and $t_i$  therefore agree with the 
geometrical three-point function $\Omega_3^{geom}$ eq. (\ref{otg}). 

\paragraph{Calculation of $C^0$.}

It remains to explicitly compute the integral $I$. Inserting $1=i\int_{i\R} dp'\ \delta(ip-ip')$ and $\delta(ip-ip')=\int_0^\infty \frac{dz}{z}\ z^{p+p'}$ yields
\bea
I &= &\int_0^\infty \frac{dz}{z}\int_{i\R}dp\ dp'\ z^{p+p'} \frac{\G(-\ell_1+p)\G(-\ell_3+p)}{\G(-\ell_1-\ell_2+p)}\G(-p)  \G(\ell_1-\ell_2-p')\G(\ell_3+1+p') \nn
\\ &=& \G(\ell_{13}^2+1) \frac{\G(-\ell_1)\G(-\ell_3)}{\G(-\ell_1-\ell_2)}\int_0^\infty \frac{dz}{z} (1+z)^{-\ell_{13}^2-1} F(-\ell_1,-\ell_3,-\ell_1-\ell_2,-z) \ .
\label{ifz}
\eea
This can be integrated with the help of the formula (\ref{izf}), yielding
\bea
I = \G(-\ell_1)\G(-\ell_3)\G(\ell_3+1) \frac{\G(\ell_{13}^2+1) \G(-\tfrac12\ell_{12}^3+1)\G(-\tfrac12 \ell_{23}^1)}{\G(1-\ell_{12}^3) \G(\tfrac12 \ell_{13}^2+1) \G(-\tfrac12 \ell_{123})} \ .
\eea
It is now easy to compute $C^0$ and compare it with the result $C^{geom}$ (\ref{cgeom}) of the geometrical calculation,
\bea
C^0 = N_1 \prod_{i=1}^3\left(N_2^{\ell_i}\frac{\G(-\ell_i)}{\G(-\ell_i-\frac12)}\right)\ C^{geom}\ ,
\eea
where $N_1,N_2$ are some normalization constants. (Such constants have been neglected in the computation.) Therefore, the $b\rar 0$ limit of the exact boundary three-point function agrees with the geometrical boundary three point function, up to an overall renormalization and a renormalization of the vertex operators.

\zeq\section{Relation with fusing matrix elements}

This section is devoted to computing certain fusing matrix elements of the $\Hp$ model, and relating them to the boundary three-point function. In the case of Liouville theory, 
the determination of the fusing matrix was used for finding the boundary three-point function \cite{pt01}. 
In the present case of the $\Hp$ model, the boundary three-point function is already known, and its relation with the fusing matrix can be deduced from the explicit formula.
Apart from testing the validity of general ideas on the structure of conformal field theories, the exercise may help address questions like: Are $AdS_2$ D-branes the only continuous, maximally symmetric D-branes in $\Hp$? How do Euclidean $AdS_2$ D-branes in $\Hp$ compare with Minkowskian $AdS_2$ D-branes in $AdS_3$? Tentative answers will be given in the Conclusion.

\subsection{An $\Hp$ fusing matrix}

The fusing matrix of the $\Hp$ model can be defined as the linear transformation between bases of $s$- and $t$-channel four-point conformal blocks. These four-point conformal blocks are supposed to be completely determined by the symmetry of the model. I will however not try to rigorously define them. Rather, I will adopt the more functional approach of using the $\Hp$-Liouville relation for deriving $s$- and $t$- channel decompositions of the boundary four-point function. I will call the objects appearing in these decompositions conformal blocks, and compute the corresponding fusing matrix. This approach will be justified a posteriori by the relation between the resulting fusing matrix elements with the boundary three-point function. However, this relation will only involve some particular combinations of fusing matrix elements; a full understanding of the $\Hp$ conformal blocks and fusing matrix is left for future work. 

I will however need one important insight from the general definition of conformal blocks based on symmetries of the model: namely, that in the $\Hp$ model the conformal blocks and fusing matrix are expected to depend on the boundary parameters $r_{ij}$. This is because the symmetry transformations of the fields (\ref{gtp}) do themselves depend on $r_{ij}$. (Like these symmetry transformations, the blocks and fusing matrix should be invariant under shifts $r_{ij}\rar r_{ij}+r_0$.)
This contrasts with the situation in say Liouville theory \cite{pt01}, where boundary parameters are purely dynamical quantities which affect neither the conformal blocks nor the fusing matrix.

\paragraph{Functional definition of the conformal blocks and fusing matrix.} 

Consider the $\nu$-basis boundary four-point function
\bea
\Omf = \la {}_{r_{41}} \Psi^{\ell_1}(\nu_1|w_1)_{r_{12}} \Psi^{\ell_2}(\nu_2|w_2)_{r_{23}} \Psi^{\ell_3}(\nu_3|w_3)_{r_{34}} \Psi^{\ell_4}(\nu_4|w_4)_{r_{41}} \ra \ .
\eea
The $s$-channel and $t$-channel four-point conformal blocks 
\bea
{\cal G}^{\ell_s}_{\lambda_{12}\lambda_{34}}\left(\ell_1 \underset{r_{12}}{|} \ell_2 \underset{r_{23}}{|} \ell_3 \underset{r_{34}}{|} \ell_4 \underset{r_{41}}{|} \right|\nu_i \left| \underset{{}_{}}{}w_i\right)
\scs 
 {\cal G}^{\ell_t}_{\lambda_{23}\lambda_{14}}\left(\ell_1 \underset{r_{12}}{|} \ell_2 \underset{r_{23}}{|} \ell_3 \underset{r_{34}}{|} \ell_4 \underset{r_{41}}{|} \right|\nu_i \left| \underset{{}_{}}{}w_i\right) \ ,
\eea
are defined as the quantities appearing in the $s$-channel and $t$-channel decompositions of $\Omf$,
\bea
\Omf &=& \sum_{\lambda_{12},\lambda_{34}}\int_{-\frac12+i\R} d\ell_s\ \left(R^L(\beta_s)\right)^{-1} C_{\lambda_{12}}{\scriptstyle \left(\ell_1\underset{r_{12}}{|} \ell_2\underset{r_{23}}{|} \ell_s \underset{r_{41}}{|}\right)} C_{\lambda_{34}}{\scriptstyle\left(\ell_3\underset{r_{34}}{|} \ell_4\underset{r_{41}}{|} \ell_s\underset{r_{23}}{|}\right)}
{\cal G}^{\ell_s}_{\lambda_{12}\lambda_{34}} , \label{sdec}
\\
&=& \sum_{\lambda_{23},\lambda_{41}}\int_{-\frac12+i\R}d\ell_t\ \left(R^L(\beta_t)\right)^{-1} 
C_{\lambda_{23}}{\scriptstyle\left(\ell_2\underset{r_{23}}{|} \ell_3\underset{r_{34}}{|} \ell_t\underset{r_{12}}{|}\right)} C_{\lambda_{41}}{\scriptstyle\left(\ell_4\underset{r_{41}}{|} \ell_1\underset{r_{12}}{|} \ell_t \underset{r_{34}}{|}\right)} {\cal G}^{\ell_t}_{\lambda_{23}\lambda_{41}},
\label{tdec}
\eea
which otherwise involve the three-point structure constant $C_\lambda$ and the $\nu$-basis reflection coefficient $ R^L_{\frac{r_{23}}{2\pi b} -\frac{i}{4b}\sgn(\nu_1+\nu_2),\frac{r_{41}}{2\pi b}+\frac{i}{4b}\sgn(\nu_1+\nu_2)}\left(b(\ell_s+1)+\frac{1}{2b}\right)$ eq. (\ref{refnu}). 

The fusing matrix is defined as realizing the change of basis between $s$- and $t$-channel blocks, 
\bea
{\cal G}^{\ell_s}_{\lambda_{12}\lambda_{34}} = \sum_{\lambda_{23},\lambda_{41}}\int_{-\frac12+i\R}d\ell_t\ 
F^{\ell_s\ell_t}_{\lambda_{12}\lambda_{34}\lambda_{23}\lambda_{14}}\!\! \left[\begin{array}{ccc} \ell_3  & {\scriptstyle r_{23}} & \ell_2
 \vspace{-1mm}
 \\ {\scriptstyle r_{34}} & & {\scriptstyle r_{12}} 
 \\ \ell_4 & {\scriptstyle r_{41}} & \ell_1
 \end{array}\right]\
{\cal G}^{\ell_t}_{\lambda_{23}\lambda_{41}} \ .
\label{gfg}
\eea
The conformal blocks and their fusion transformation will be depicted as
\bea
\psset{unit=.35cm}
\pspicture[](-7,-3)(5,3)
\psline(3,2)(2,0)(-2,0)(-3,2)\psline(-2,0)(-3,-2)\psline(2,0)(3,-2)
\rput(3.5,3){$\ell_2$} \rput(3.5,-3){$\ell_{1}$} \rput(-3.5,3){$\ell_3$} \rput(-3.5,-3){$\ell_{4}$}
\rput(0,3){$r_{23}$} \rput(0,-3){$r_{41}$} \rput(3.5,0){$r_{12}$} \rput(-3.5,0){$r_{34}$}
\rput(0,.7){$\ell_{s}$} \rput(1.5,-.7){$\lambda_{12}$}\rput(-1.5,-.7){$\lambda_{34}$}
\endpspicture
\overset{\displaystyle F^{\ell_s\ell_t}}{\longrightarrow}
\pspicture[](-5,-3.5)(5,3.5)
\psline(2,3)(0,2)(0,-2)(2,-3)\psline(0,-2)(-2,-3)\psline(0,2)(-2,3)
\rput(3,3.5){$\ell_2$}\rput(-3,3.5){$\ell_{3}$} \rput(3,-3.5){$\ell_{1}$}\rput(-3,-3.5){$\ell_{4}$}
\rput(3,0){$r_{12}$} \rput(-3,0){$r_{34}$} \rput(0,3.5){$r_{23}$} \rput(0,-3.5){$r_{41}$}
\rput(.7,0){$\ell_t$} \rput(-.8,1.5){$\lambda_{23}$}\rput(-.8,-1.5){$\lambda_{41}$}
\endpspicture
\ \ .
\label{fpict}
\eea

\paragraph{$\Hp$ conformal blocks from Liouville conformal blocks.}

The $\Hp$ boundary four-point function can be written in terms of a Liouville boundary six-point function as 
\cite{hr06}:
\bea
 \Omf = \delta(\tsum\nu_i) |\tsum\nu_i w_i| \left|\frac{y_{12}\prod_{i<i'} w_{ii'}} {\prod_{a=1,2}\prod_i (y_a-w_i)}\right|^\frac{1}{2b^2}  \la \prod_{i=1}^4 B^{\beta_i}(w_i)\ \  B^{-\frac{1}{2b}}(y_1) B^{-\frac{1}{2b}}(y_2)\ra\ ,
\label{omf}
\eea
where $\beta_i=b(\ell_i+1)+\tfrac{1}{2b}$ as before, the Liouville boundary parameter is still given by eq. (\ref{srphi}), and $y_1,y_2$ are still defined as  the zeroes of a function $\varphi(y)$ (\ref{vpy}). 
The idea is now to decompose the Liouville six-point function in terms of Liouville structure constants and conformal blocks, out of which the $\Hp$ structure constants $C_\lambda$ and conformal blocks should be reconstructed. The details of the decomposition are quite sensitive on signs of the isospin variables $\nu_i$, which determine the positions of the Liouville degenerate fields $B^{-\frac{1}{2b}}(y_1),B^{-\frac{1}{2b}}(y_2)$ on the worldsheet boundary. (In some cases, the degenerate fields can even live in the bulk.) Such subtleties would be very relevant to a rigorous definition of the conformal blocks; but here I will neglect them and assume
\bea
(\sgn\nu_1,\sgn\nu_2,\sgn\nu_3,\sgn\nu_4) = (+,-,-,-) \ \ \Rightarrow \ \ w_2<y_1<w_3<y_2<w_4\ .
\eea
Now I claim that, in this regime, $s$-channel blocks can be built in terms of Liouville blocks as 
\bea
{\cal G}^{\ell_s}_{\lambda_{12}\lambda_{34}} = N^+_{r_{41},r_{23}} (\ell_s) \sum_{\eta_2,\eta_4} T^{[+1(2)s]}_{\lambda_{12},\eta_2} T^{[+s3(4)]}_{\lambda_{34},\eta_4} 
\psset{unit=.35cm}
\pspicture[](-4,-2.6)(4,2.2)
\psline(3,2)(2,0)(-2,0)(-3,2)
\psline(-3,-2)(-2,0) \psline(3,-2)(2,0)
\thecoil(2.2,.4)(1.6,2.2) \thecoil(-2.2,-.4)(-3.2,1)
\rput(2.8,-0.2){$\eta_2$} \rput(-1.6,-.7){$\eta_4$}
\rput(3.5,2.2){$2$} \rput(-3.5,2.2){$3$} \rput(-3.5,-2.2){$4$} \rput(3.5,-2.2){$1$}
\rput(0,.6){$s$}
\endpspicture
\ , 
\label{gls}
\eea
where, in the diagrammatic representation of the standard six-point Liouville blocks, the wiggly lines are the degenerate fields, whose fusion channels are labelled $\eta=\pm$ like in the four-point Liouville blocks of eq. (\ref{fldeg}), and the solid lines are the generic fields with momenta $\beta_1,\beta_2,\beta_3,\beta_4,\beta_s$.  (The prefactors in eq. (\ref{omf}) are implicitly included in the Liouville blocks.) (Remember that $N^\sigma_{r,r'}(\ell)$ was defined in (\ref{rrr}), and $T_{\lambda,\eta}^{[\sgn\nu_i]}$ in (\ref{ctf}).)

The proof that such $s$-channel blocks do indeed satisfy eq. (\ref{sdec}) is straightforward, given the relation (\ref{ccct}) between Liouville and $\Hp$ boundary three-point structure constants. It is of course also possible to find $t$-channel blocks satisfying eq. (\ref{tdec}),
\bea
{\cal G}^{\ell_t}_{\lambda_{23}\lambda_{41}}=N^-_{r_{12},r_{34}}(\ell_t) \sum_{\eta_2,\eta_4}
T^{[+t(2)3]}_{\lambda_{23},\eta_2} T^{[+1t(4)]}_{\lambda_{41},\eta_4} 
\psset{unit=.3cm}
\pspicture[](-4,-3)(4,3)
\psline(2,3)(0,2)(0,-2)(2,-3)
\psline(-2,-3)(0,-2)\psline(-2,3)(0,2)
\thecoil(.4,2.2)(-1,3.2) \thecoil(-.4,-2.2)(-2.2,-1.2)
\rput(.8,1.5){$\eta_2$} \rput(0,-2.9){$\eta_4$}
\rput(2.5,3.2){$2$} \rput(2.5,-3.2){$1$} \rput(-2.5,-3.2){$4$} \rput(-2.5,3.2){$3$}
\rput(.6,0){$t$}
\endpspicture
\label{glt}
\eea
Let me now derive the fusing matrix which relates these $s$- and $t$-channel blocks.

\paragraph{$\Hp$ fusing matrix from Liouville fusing matrix.} 

The relation between the Liouville conformal blocks appearing in the $\Hp$ $s$- and $t$-channel conformal blocks is given by the Liouville fusing matrix, which is defined by \cite{tes01a}
\bea
\psset{unit=.3cm}
\pspicture[](-4,-1.8)(4,1.8)
\psline(3,2)(2,0)(-2,0)(-3,2)
\psline(-3,-2)(-2,0) \psline(3,-2)(2,0)
\rput(3.5,2.2){$2$} \rput(-3.5,2.2){$3$} \rput(-3.5,-2.2){$4$} \rput(3.5,-2.2){$1$}
\rput(0,.6){$s$}
\endpspicture
= 
\int_{\frac{Q}{2}+i\R} d\beta_t\ F^L_{\beta_s\beta_t}\!\!\left[\begin{array}{cc}\beta_3 & \beta_2 \\ \beta_4 &\beta_1 \end{array}\right]\ 
\psset{unit=.3cm}
\pspicture[](-4,-3.2)(4,3)
\psline(2,3)(0,2)(0,-2)(2,-3)
\psline(-2,-3)(0,-2)\psline(-2,3)(0,2)
\rput(2.5,3.2){$2$} \rput(2.5,-3.2){$1$} \rput(-2.5,-3.2){$4$} \rput(-2.5,3.2){$3$}
\rput(.6,0){$t$}
\endpspicture
\ .
\label{fliou}
\eea
Applying this relation to the Liouville blocks appearing in the formulas for $\Hp$ four-point blocks (\ref{gls}) and (\ref{glt}) yields an $\Hp$ fusing matrix satisfying eq. (\ref{gfg}):
\begin{multline}
F^{\ell_s\ell_t}_{\lambda_{12}\lambda_{34}\lambda_{23}\lambda_{14}} \!\!\left[\begin{array}{ccc} \ell_3  & {\scriptstyle r_{23}} & \ell_2
 \vspace{-1mm}
 \\ {\scriptstyle r_{34}} & & {\scriptstyle r_{12}} 
 \\ \ell_4 & {\scriptstyle r_{41}} & \ell_1
 \end{array}\right] = \frac{N_{r_{41},r_{23}}^+(\ell_s)}{N^-_{r_{12},r_{34}}(\ell_t)} 
 \\ \times 
 \sum_{\eta_2,\eta_4} T^{[+1(2)s]}_{\lambda_{12},\eta_2} T^{[+s3(4)]}_{\lambda_{34},\eta_4}
F^L_{\beta_s\beta_t} \!\!\left[\begin{array}{cc}\beta_3 & \beta_2-\tfrac{\eta_2}{2b} \\ \beta_4-\tfrac{\eta_4}{2b} &\beta_1\end{array}\right] \left(T^{-1}\right)^{[+t(2)3]}_{\eta_2,\lambda_{23}} \left(T^{-1}\right)^{[+1t(4)]}_{\eta_4,\lambda_{41}} \ .
\label{fmm}
\end{multline}
Notice that the four
Liouville fusing matrix elements appearing in this formula are not all independent, but can be related to any two of them via linear equations whose coefficients are products of Gamma functions. (See Appendix \ref{apfl}.) 

It can actually be proved that this fusing matrix satisfies a Pentagon equation, but this is outside the scope of this article. In general conformal field theories, the Pentagon equation is the structural reason for the existence of a relation between the fusing matrix and the boundary three-point function. Here I will however derive such a relation by direct calculation.

\subsection{Discrete representations of $\USLR$\label{subrep}}

This subsection is a technical interlude devoted to the definition and study of the discrete representations of $\USLR$. There may seem to be no physical motivation for studying such representations in the context of the $\Hp$ model, whose spectrum is purely continuous. However, it will turn out that discrete representations play a crucial role in the relation between the fusing matrix and the boundary three-point function.\footnote{Note that by focusing on the $\USLR$ horizontal subgroup I am still ignoring the rest of the infinite-dimensional symmetry group of the model. 
Representations of $\USLR$ can however easily be extended to highest-weight representations of the full symmetry group. Anyway, since discrete representations are absent from the spectrum, their structure will be of no importance in the following. Only formal properties like the allowed values of the spins will be needed.}

\paragraph{Discrete representations and discrete fields.}

There are two series of discrete representations, called $D^+_\ell$ and $D^-_\ell$. A representation $D^\pm_\ell$ is defined as having a state which is annihilated by the generator $J^\mp$ of the $s\ell_2$ Lie algebra, whose commutation relations and quadratic Casimir operator are
\bea
[J^3,J^\pm]=\pm J^\pm,\ [J^+,J^-]=-2J^3,\quad C= -(J^3)^2 +\tfrac12(J^+J^-+J^-J^+)\ .
\eea
The eigenvalues of $C$ are labelled in terms of the spin $\ell$ as $C=-\ell(\ell+1)$, and the eigenvalues of $J^3$ are called $m$. The $J^-$-annihilated state of a $D^+_\ell$ representation can have either $m=\ell+1$ or $m=-\ell$. In the case $\ell\in\frac12 \Z$, such a state must have $m>0$, otherwise a $J^+$-annihilated state appears at $J^3=-m$, and the representation is finite-dimensional instead of being discrete. In the case of generic $\ell$ however, both $D^+_\ell$ representations should be accepted, but distinguishing them will not matter in the following. I will also ignore the special case $\ell\in\frac12 \Z$. Note however that discrete representations of $\SLR$ must have $\ell\in \frac12 \Z$, whereas discrete representations of the universal cover $\USLR$ exist for all $\ell\in \C$.

A field $\Psi^\ell(t)$ belonging to the $D^\pm_\ell$ representation can be analytically continued to  the half-plane $U^\pm\equiv \{\pm \Im t > 0\}$ \cite{ggv66}.  So if $\Psi^{\ell_2}(t_2)\in D^\sigma_{\ell_2}$ with $\sigma=\pm$, then 
the $t$-basis three-point function $\Omega_3$ (\ref{omt}) must be analytic in $t_2\in U^\pm$. This constrains its behaviours near $t_2=t_1$ and $t_2=t_3$. For instance, near $t_2=t_1$ the relevant factors of $\Omega_3$ behave as 
$\Omega_3\propto |t_{12}|^{\ell_{12}^3} e^{\frac{k-2}{2}r_{12}\sgn t_{12}} C_{-\sgn t_{12}}$, which has an analytic continuation to $t_2\in U^\sigma$ provided $e^{i\pi \sigma \ell_{12}^3} e^{-(k-2)r_{12}} C_- = C_+$.
Together with the condition $e^{i\pi \sigma \ell_{23}^1} e^{(k-2)r_{23}} C_+=C_-$ from $t_2\sim t_3$, this is equivalent to
\bea
\ell_2 &\in & \sigma \frac{k-2}{2\pi i}(r_{12}-r_{23}) +\Z\ ,
\label{ldz}
\\
C_\lambda &=& \lambda^n\ e^{\frac{\lambda}{2}\left[i\pi \sigma (\ell_1-\ell_3)-\frac{k-2}{2}(r_{12}+r_{23})\right]} C_0\ ,
\label{clco}
\eea
where $C_0$ is a $\lambda$-independent constant, and $n\in\{0,1\}$ is the parity of the element of $\Z$ above. The condition on $\ell_2$ depends only on the field ${}_{r_{12}}\Psi^{\ell_2}(t_2)_{r_{23}}$ and not on the other fields in the three-point function, and it is the condition for that field to be discrete.

The interesting feature of discrete representations is therefore the disappearance of the multiplicity $\lambda$ in the boundary interactions: a three-point function involving a discrete representation is determined in terms of only one structure constant $C_0$, instead of $C_\pm$ in the generic case. 

\paragraph{Discrete $\nu$-basis fields.} 

Since the investigation of the fusing matrix in $\Hp$ heavily relied on the $\nu$-basis, it will be necessary to understand how fields transforming in discrete representations behave in the $\nu$-basis. 
The analyticity of discrete fields for $t\in U^\pm$ translates into corresponding $\nu$-basis fields  
$\Psi^\ell(\nu)=|\nu|^{\ell+1}\int_\R dt\ e^{i\nu t} \Psi^\ell(t)$ vanishing for $\pm\nu>0$. How does this simplify the coefficients $T^{[\pm i(j)k]}_{\lambda,\eta}$ eq. (\ref{tlp})-(\ref{tlm}), which enter the formula for the fusing matrix? The coefficients $T^{[\sigma 1(2)3]}_{\lambda,\eta}$ are defined for $\sgn\nu_2=-\sigma$, and the explicit formula shows
\bea
\ell_2 \in -\sigma \frac{k-2}{2\pi i}(r_{12}-r_{23}) +\Z \quad \Rightarrow\quad 
T^{[\sigma 1(2)3]}_{\lambda,-}=0\ .
\label{sfs}
\eea
What if it is the third field in $\Omega_3$ which belongs to a discrete representation $D^\sigma_{\ell_3}$? 
Then similarly $T^{[\sigma 12(3)]}_{\lambda,-}=0$, and the relation (\ref{tmt}) yields 
\bea
\ell_3\in \sigma \frac{k-2}{2\pi i}(r_{23}-r_{31}) +\Z \quad \Rightarrow\quad
\frac{T^{[\sigma 1(2)3]}_{\lambda,+}}{T^{[\sigma 1(2)3]}_{\lambda,-}} = \frac{M^{(23)1}_{++}}{M^{(23)1}_{-+}} \ ,
\label{tfs}
\eea
so that $T^{[\sigma 1(2)3]}_{\lambda,+}$ must have the same $\lambda$-dependence as $T^{[\sigma 1(2)3]}_{\lambda,-}$. Finally, what if it is the first field in $\Omega_3$ which now belongs to $D^\sigma_{\ell_1}$? Just use the explicit formulas for $T^{[\sigma 1(2)3]}_{\lambda,\eta}$ to read off how they behave under $1\lrar 3$, and deduce from the previous case 
\bea
\ell_1\in -\sigma \frac{k-2}{2\pi i}(r_{31}-r_{12}) +\Z \quad \Rightarrow\quad 
\frac{T^{[\sigma 1(2)3]}_{\lambda,+}}{T^{[\sigma 1(2)3]}_{\lambda,-}} = \frac{M^{(21)3}_{++}}{M^{(21)3}_{-+}}\ \frac{\sin\left(\pi \ell_2-i\sigma \frac{r_{23}-r_{12}}{2b^2}\right)}{\sin\left(\pi \ell_2+i\sigma \frac{r_{23}-r_{12}}{2b^2}\right)} \ .
\label{ffs}
\eea

\subsection{Relation fusing matrix -- boundary three-point function}

\paragraph{The case of Liouville theory.} 

Let me begin with recalling the form of this relation in Liouville theory. On the one hand this will be useful in the derivation of the $\Hp$ relation, on the other hand this will illustrate what type of relation should be expected. 

The Liouville boundary three-point function (\ref{clbbb}) is related to the Liouville fusing matrix (\ref{fliou}) by \cite{pt01}
\begin{multline}
C^L\left(\beta_1\underset{s_{12}}{|}\beta_2 \underset{s_{23}}{|} \beta_3 \underset{s_{31}}{|}\right) 
\\
=
R^L_{s_{31},s_{12}}(\beta_1) \frac{g^L_{s_{31},s_{12}}(\beta_1)}{g^L_{s_{12},s_{23}}(\beta_2) g^L_{s_{23},s_{31}}(\beta_3)} 
F^L_{\frac{Q}{2}+is_{23},\beta_1}\!\!\left[\begin{array}{cc}\beta_3 & \beta_2 \\ \frac{Q}{2}+is_{31} & \frac{Q}{2}+is_{12} \end{array}\right] \ , 
\label{flcl}
\end{multline}
where the function $g^L_{s,s'}(\beta)$, which may be seen as a sort of square root of the reflection coefficient (\ref{rlss}) and satisfies $g^L_{s,s'}(\beta)=R^L_{s,s'}(Q-\beta) g^L_{s,s'}(Q-\beta)$, is
\bea
g^L_{s,s'}(\beta) &=&  \mu_L^{\frac12 b^{-1}\beta} \ \frac{\G_b(Q) \G_b(Q-2\beta) \G_b(Q+2is) \G_b(Q-2is')}{\prod_{\pm\pm} \G_b(Q-\beta\pm is \pm is')} \ .
\eea
The basic idea, which is originally due to Cardy \cite{car89}, is therefore to associate some momenta $\beta_{ij}=\frac{Q}{2}+is_{ij}$ to the boundary conditions $s_{ij}$. These momenta are then used 
as inputs in the fusing matrix \cite{run98}. 

\paragraph{Peculiarities of the $\Hp$ model.}

Unlike Liouville theory, the $\Hp$ model does not a priori conform to the assumptions which would make these ideas work. In particular, the $\SLC$ representations appearing in the bulk spectrum are labelled by their sole spin, whereas the $\USLR$ representations appearing in the boundary spectrum are labelled by a spin and an extra continuous parameter $\alpha=r-r'$ depending on the boundary parameters $r,r'$. Associating bulk spins to the boundary conditions may be useful to some extent for understanding the moduli space of D-branes in $\Hp$ \cite{rib05b}, but the inputs in the $\Hp$ fusing matrix rather need to be pairs $(\ell,\alpha)$ as in the boundary spectrum.

Another feature of the $\Hp$ case is the presence of a multiplicity index $\lambda$ in the three-point structure constant $C_\lambda$, and of four corresponding indices in the fusing matrix. The generic expectation \cite{bppz99,frs04} is that such multiplicities should also appear as indices of the boundary fields themselves. There should indeed be a correspondence between boundary fields and three-point vertices: 
\bea
\psset{unit=.35cm}
\pspicture[](-3,-3)(5,3)
\psline(-3,0)(3,0)
\psdots[dotscale=2,dotangle=45,dotstyle=+](0,0)
\rput(0,1){$\Psi^{\ell_2}$} 
\rput(-1.5,-.7){$r_{12}$}
\rput(1.5,-.7){$r_{23}$}
\endpspicture
\pspicture[](-2,0)(2,0)
\psline[linewidth=2pt,arrows=->](-1.5,0)(1.5,0)
\endpspicture
\pspicture[](-6,-3)(3,3)
\psline(0,0)(0,2) \rput(0,3){$\ell_2$}
\rput{-120}(0,0){\psline(0,0)(0,2) \rput{*0}(0,3){$\ell_{23}$}}
\rput{120}(0,0){\psline(0,0)(0,2) \rput{*0}(0,3){$\ell_{12}$}}
\rput{-60}(0,0){\rput{*0}(0,3){$r_{23}$}}
\rput{60}(0,0){\rput{*0}(0,3){$r_{12}$}}
\rput(0,-3){$r_0$} \rput(0,-1){$\lambda_2$}
\endpspicture
\eea
Which spins $\ell_{12},\ell_{23}$ should correspond to the boundary conditions $r_{12},r_{23}$? What should $r_0$ and $\lambda_2$ be? The idea proposed here is to choose $\ell_{ij}$ as discrete spins, which would eliminate the index $\lambda_2$ as explained in the previous subsection, and determine $r_0$. The relation between boundary three-point function and the fusing matrix will then be of the type:
\bea
\psset{unit=.35cm}
\pspicture[](-3,-3)(5,3)
\psline(0,0)(0,2) \rput(0,3){$\ell_2$}
\rput{-120}(0,0){\psline(0,0)(0,2) \rput{*0}(0,3){$\ell_{1}$}}
\rput{120}(0,0){\psline(0,0)(0,2) \rput{*0}(0,3){$\ell_{3}$}}
\rput{-60}(0,0){\rput{*0}(0,3){$r_{12}$}}
\rput{60}(0,0){\rput{*0}(0,3){$r_{23}$}}
\rput(0,-3){$r_{31}$} \rput(0,-1){$\lambda$}
\endpspicture
\pspicture[](-2,0)(2,0)
\psline[linewidth=2pt,arrows=->](-1.5,0)(1.5,0)
\endpspicture
\pspicture[](-7,-3)(5,3)
\psline(3,2)(2,0)(-2,0)(-3,2)\psline(-2,0)(-3,-2)\psline(2,0)(3,-2)
\rput(3.5,3){$\ell_2$} \rput(3.5,-3){$\ell_{12}$} \rput(-3.5,3){$\ell_3$} \rput(-3.5,-3){$\ell_{31}$}
\rput(0,3){$r_{23}$} \rput(0,-3){$r_0$} \rput(3.5,0){$r_{12}$} \rput(-3.5,0){$r_{31}$}
\rput(0,1){$\ell_{23}$} \rput(1.5,-.7){$\lambda_2$}\rput(-1.5,-.7){$\lambda_3$}
\endpspicture
\overset{\displaystyle F^{\ell_{23}\ell_1}}{\longrightarrow}
\pspicture[](-5,-3.5)(5,3.5)
\psline(2,3)(0,2)(0,-2)(2,-3)\psline(0,-2)(-2,-3)\psline(0,2)(-2,3)
\rput(3,3.5){$\ell_2$}\rput(-3,3.5){$\ell_{3}$} \rput(3,-3.5){$\ell_{12}$}\rput(-3,-3.5){$\ell_{31}$}
\rput(3,0){$r_{12}$} \rput(-3,0){$r_{31}$} \rput(0,3.5){$r_{23}$} \rput(0,-3.5){$r_0$}
\rput(1,0){$\ell_1$} \rput(-.7,1.5){$\lambda$}\rput(-.7,-1.5){$\lambda_1$}
\endpspicture
\ ,
\label{vof}
\eea
where the dependence of the fusing matrix $F^{\ell_{23}\ell_1}$ on $\lambda_1,\lambda_2,\lambda_3$ is trivial thanks to the spins $\ell_{ij}$ being discrete.

\paragraph{Derivation of the relation by direct calculation.}

I will not seek further guidance from general structural ideas, but rather from the explicit formulas. Namely, I will use the relations between the $\Hp$ and Liouville three-point structure constants (\ref{ccct}), then between the Liouville structure constant and fusing matrix (\ref{flcl}), and finally between the Liouville and $\Hp$ fusing matrices (\ref{fmm}). Specifically, start with
\bea
C_\lambda = \sum_{\eta=\pm} 
C^L_\sigma(\beta_1|\beta_2-\tfrac{\eta}{2b}|\beta_3)\  C^L_\sigma(\beta_2|-\tfrac{1}{2b}|Q-\beta_2+\tfrac{\eta}{2b})\
\left(T^{-1}\right)^{[\sigma 1(2)3]}_{\eta,\lambda}\ ,
\eea
and insert the expression for $C^L_\sigma(\beta_1|\beta_2-\tfrac{\eta}{2b}|\beta_3)$ in terms of the Liouville fusing matrix in the case $\sigma=+$, 
\begin{multline}
C_\lambda =R^L_{\frac{r_{31}}{2\pi b}+\frac{i}{4b},\frac{r_{12}}{2\pi b}-\frac{i}{4b}}(\beta_1) \frac{g^L_{\frac{r_{31}}{2\pi b}+\frac{i}{4b},\frac{r_{12}}{2\pi b}- \frac{i}{4b}}(\beta_1)}{g^L_{\frac{r_{23}}{2\pi b}-\frac{i}{4b},\frac{r_{31}}{2\pi b}+\frac{i}{4b}}(\beta_3)} 
\sum_{\eta=\pm}
\frac{C^L_+(\beta_2|-\tfrac{1}{2b}|Q-\beta_2+\tfrac{\eta}{2b})
}{ g^L_{\frac{r_{12}}{2\pi b}-\frac{i}{4b},\frac{r_{23}}{2\pi b}- \frac{i}{4b}}(\beta_2-\tfrac{\eta}{2b})} 
\\ \times 
\left(T^{-1}\right)^{[+ 1(2)3]}_{\eta,\lambda}
F^L_{\frac{Q}{2}-\frac{r_{23}}{2\pi i b}+\frac{1}{4b},\beta_1}\!\!\left[\begin{array}{ccc}\beta_3 & & \beta_2-\tfrac{\eta}{2b} \\ \frac{Q}{2}-\frac{r_{31}}{2\pi ib}-\frac{1}{4b} & \hspace{2mm}& \frac{Q}{2}-\frac{r_{12}}{2\pi ib}+\frac{1}{4b} \end{array}\right] \ .
\label{chfl}
\end{multline}
This combination $\sum_{\eta=\pm}$ of two $F^L$ matrices should be compared to the combination appearing in the following rewriting of the $\Hp$ fusing matrix (\ref{fmm}), where I use the property $T^{[+12(3)]}=T^{[-1(3)2]}$:  
\begin{multline}
 \sum_{\lambda_1} T_{\lambda_1,\eta_0}^{[-,12,(31),1]}
 F^{\ell_{23}\ell_1}_{\lambda_{2}\lambda_{3}\lambda\lambda_1} \!\!\left[\begin{array}{ccc} \ell_3  & {\scriptstyle r_{23}} & \ell_2
 \vspace{-1mm}
 \\ {\scriptstyle r_{31}} & & {\scriptstyle r_{12}} 
 \\ \ell_{31} & {\scriptstyle r_0} & \ell_{12}
 \end{array}\right] 
 \\
=  \frac{N^{+}_{r_0,r_{23}}(\ell_{23})}{N^{-}_{r_{12},r_{31}}(\ell_1)} T^{[-,23,(31),3]}_{\lambda_3,\eta_0}\sum_{\eta=\pm} T^{[+,12,(2),23]}_{\lambda_2,\eta} \left(T^{-1}\right)^{[+1(2)3]}_{\eta,\lambda} F^L_{\beta_{23},\beta_1}\!\!\left[\begin{array}{cc} \beta_3 & \beta_2-\tfrac{\eta}{2b} \\ \beta_{31}-\tfrac{\eta_0}{2b} & \beta_{12}\end{array}\right]\ .
\label{fhfl}
\end{multline}
The $F^L$ fusing matrices which appear in the last two equations are equal provided their arguments are identical modulo reflection $\beta\rar Q-\beta$. This is the case if one assumes $\eta_0=+$ and $\beta_{ij}=b(\ell_{ij}+1)+\frac{1}{2b}$ with  
\bea
\ell_{ij}=-\frac12-\frac{r_{ij}}{2\pi i b^2}+\frac{1}{4b^2}\ .
\label{lr}
\eea
This relation between spins and boundary parameters agrees with the one proposed in \cite{rib05b}. However,
the idea is now to interpret the corresponding representations as discrete representations. This is possible if the relation $\ell_{ij}\in -\frac{k-2}{2\pi i}(r_{ij}-r_0)+\Z$ is obeyed. And this relation indeed holds provided the following assumption is made: 
\bea
r_0 = i\pi \left(\frac12-b^2\right)\ . 
\label{rz}
\eea
Then, according to the formulas (\ref{tfs}) and (\ref{ffs}), 
the factors $T_{\lambda_1,\eta_0}^{[-,12,(31),1]}$, $T^{[-,23,(31),3]}_{\lambda_3,\eta_0}$ and $T^{[+,12,(2),23]}_{\lambda_2,\eta}$ simplify (without vanishing), in the sense that  
their $\lambda$ and $\eta$-dependences disentangle. In particular, the $\lambda_2$-dependence in eq. (\ref{fhfl}) can be rewritten as a prefactor, outside the sum $\sum_{\eta=\pm}$.

\paragraph{Test and results.}

Now that the parameters $r_0,\ell_{ij}$ are fixed, comes the test: are the combinations of two $F^L$-matrices in (\ref{chfl}) and (\ref{fhfl}) proportional up to an overall factor? Direct calculations (which use eq. (\ref{tfs})) indeed show that they are, thanks to the following identity, valid for any $\sigma=\pm$:
\bea
\frac{1}{R^L_{\frac{r_{12}}{2\pi b}-\sigma \frac{i}{4b},\frac{r_{23}}{2\pi b}+\sigma \frac{i}{4b}}(\beta_2)}
\frac{g^L_{\frac{r_{12}}{2\pi b}-\sigma \frac{i}{4b},\frac{r_{23}}{2\pi b}-\sigma \frac{i}{4b}}(Q-\beta_2-\tfrac{1}{2b})}{g^L_{\frac{r_{12}}{2\pi b}-\sigma \frac{i}{4b},\frac{r_{23}}{2\pi b}-\sigma \frac{i}{4b}}(\beta_2-\tfrac{1}{2b})}
=\frac{T^{[\sigma,12,(2),23]}_{\lambda_2,+}}{T^{[\sigma,12,(2),23]}_{\lambda_2,-}}\ .
\eea
It is then possible to define coefficients of the type
\bea
g^\lambda_{r,r'}(\ell) = e^{-\lambda\left[i\frac{\pi}{2}(\ell+1)+\frac{r+r'}{4b^2}\right]} g^0_{r,r'}(\ell)\ ,
\eea
where $g^0_{r,r'}(\ell)$ is $\lambda$-independent, such that for all $\lambda_2,\lambda_3=\pm$ 
\bea
C_\lambda\left(\ell_1\underset{r_{12}}{|} \ell_2\underset{r_{23}}{|} \ell_3\underset{r_{31}}{|}\right) = R_{r_{31},r_{12}}(\ell_1) \sum_{\lambda_1=\pm}\frac{g^{\lambda_1}_{ r_{31}, r_{12}}(\ell_1)}{g^{\lambda_2}_{ r_{12}, r_{23}}(\ell_2) g^{\lambda_3}_{ r_{23}, r_{31}}(\ell_3)} F^{\ell_{23}\ell_1}_{\lambda_{2}\lambda_{3}\lambda\lambda_1} \!\!\left[\begin{array}{ccc} \ell_3  & {\scriptstyle r_{23}} & \ell_2
 \vspace{-1mm}
 \\ {\scriptstyle r_{31}} & & {\scriptstyle r_{12}} 
 \\ \ell_{31} & {\scriptstyle r_0} & \ell_{12}
 \end{array}\right] \ .
\label{chfh}
\eea
This is the sought-after expression for the boundary three-point function in terms of fusing matrix elements, which depend on the particular arguments $r_0$ and $\ell_{ij}$ defined above. 
This result can be  rewritten in terms of a ``partly discrete fusing matrix'' $\tilde{F}$ such that
\bea
C_\lambda\left(\ell_1\underset{r_{12}}{|} \ell_2\underset{r_{23}}{|} \ell_3\underset{r_{31}}{|}\right) = R_{r_{31},r_{12}}(\ell_1) \frac{g^{0}_{ r_{31}, r_{12}}(\ell_1)}{g^{0}_{ r_{12}, r_{23}}(\ell_2) g^{0}_{ r_{23}, r_{31}}(\ell_3)} \tilde{F}^{\ell_{23}\ell_1}_\lambda \!\!\left[\begin{array}{cc} \ell_3 & \ell_2 \\ \ell_{31} & \ell_{12} \end{array}\right]\ .
\eea
In this notation, the $\Hp$ result becomes very similar to the Liouville result (\ref{flcl}).

\paragraph{Representation-theoretic discussion.}

Let me now check that the use of discrete representations in the fusing matrix, as suggested by the above calculations, is actually compatible with the algebraic properties of these representations. Unfortunately, the fusion products of vertex operators with $\USLR$ symmetry, and even the tensor products of $\USLR$ representations, are apparently unknown. However, some features can be extrapolated from the known $\SLR$ representations, where tensor products of the type $D^+\otimes D^-$ are expected to yield continuous representations (and possibly discrete ones), whereas tensor products $D^+\otimes D^+$ or $D^-\otimes D^-$ only yield discrete representations. These statements should also hold for fusion products of $\USLR$ representations. 

It is therefore important to determine whether the discrete representations of spins $\ell_{ij}$ (\ref{lr}) belong to the $D^+$ or to the $D^-$ series. According to the rule (\ref{ldz}), and taking good care of the orientation of the worldsheet boundary, the discrete representations are found to be $D^+_{\ell_{12}}$, $D^-_{\ell_{31}}$ and $D^\pm_{\ell_{23}}$. The sign in $D^\pm_{\ell_{23}}$ depends on a choice of orientation, as can be seen in the following oriented depiction of the fusing matrix (\ref{vof}),
\bea
\psset{unit=.25cm}
\pspicture[](-7,-3)(5,3)
\psline(3,2)(2,0)(-2,0)(-3,2)\psline(-2,0)(-3,-2)\psline(2,0)(3,-2)
\psline[arrowsize=2pt 4,arrowinset=.6]{->}(3,-2)(2.3,-.6)
\psline[arrowsize=2pt 4,arrowinset=.6]{->}(1,0)(-.4,0)
\psline[arrowsize=2pt 4,arrowinset=.6]{->}(-2,0)(-2.7,-1.4)
\rput(3.5,3){$\ell_2$} \rput(3.5,-3){$\ell_{12}$} \rput(-3.5,3){$\ell_3$} \rput(-3.5,-3){$\ell_{31}$}
%\rput(0,3){$r_{23}$} \rput(0,-3){$r_0$} \rput(3.5,0){$r_{12}$} \rput(-3.5,0){$r_{31}$}
\rput(0,1.5){$\ell_{23}$} 
%\rput(1.5,-.7){$\lambda_2$}\rput(-1.5,-.7){$\lambda_3$}
\endpspicture
\overset{\displaystyle F^{\ell_{23}\ell_1}}{\longrightarrow}
\pspicture[](-5,-3.5)(5,3.5)
\psline(2,3)(0,2)(0,-2)(2,-3)\psline(0,2)(-2,3)\psline(0,-2)(-2,-3)
\psline[arrowsize=2pt 4,arrowinset=.6]{->}(2,-3)(.6,-2.3)
\psline[arrowsize=2pt 4,arrowinset=.6]{->}(0,-2)(-1.4,-2.7)
\rput(3,3.5){$\ell_2$}\rput(-3,3.5){$\ell_{3}$} \rput(3,-3.5){$\ell_{12}$}\rput(-3,-3.5){$\ell_{31}$}
%\rput(3,0){$r_{12}$} \rput(-3,0){$r_{31}$} \rput(0,3.5){$r_{23}$} \rput(0,-3.5){$r_0$}
\rput(1,0){$\ell_1$} 
%\rput(-.7,1.5){$\lambda$}\rput(-.7,-1.5){$\lambda_1$}
\endpspicture
\ .
\eea
In this picture, incoming arrows denote $D^+$ representations, outgoing arrows denote $D^-$ representations, and lines without arrows denote $C$ (Continuous) representations. The vertices involving discrete representations are all of the 
$ \psset{unit=.23cm}
\pspicture[](-2,-1)(2,2) 
\psline(0,0)(0,2) \rput{-120}(0,0){\psline(0,0)(0,2) \psline[arrowsize=2pt 4,arrowinset=.6]{->}(0,2)(0,.6)} \rput{120}(0,0){\psline(0,0)(0,2) \psline[arrowsize=2pt 4,arrowinset=.6]{->}(0,0)(0,1.4)} 
\endpspicture $
type, and they therefore correspond to non-vanishing $D^+\otimes D^-\rar C$ intertwiners. 

\zeq\section{Conclusions and speculations}

\paragraph{Another limit of the boundary three-point function.}

The geometrical (or minisuperspace) limit of the boundary three-point function has provided a non-trivial check of the exact formula, see subsection \ref{subgeom}. In this limit, the brane parameters $r_{12},r_{23},r_{31}$ are kept fixed, and the limit then exists only provided they are all equal. It is however interesting to consider another $b\rar 0$ limit, where the quantities $R_{ij}\equiv\frac{r_{ij}}{2\pi b^2}$ are kept fixed. 
This limit no longer requires them to be equal, and can be explicitly computed from eqs. (\ref{omt}) and (\ref{fef}):
\begin{multline}
\Omega_3\ \sim \ |t_{12}|^{\ell_{12}^3}|t_{13}|^{\ell_{13}^2} |t_{23}|^{\ell_{23}^1} e^{\pi R_{12}\sgn t_{12} +\pi R_{23}\sgn t_{23} +\pi R_{31}\sgn t_{31}} 
\\
\frac{\G(-\ell_{13}^2) \G(-\ell_{12}^3) \G(-\ell_{123}-1)}{\prod_{i=1}^3 \G(-2\ell_i-1)} 
\sum_{\sigma=\pm} e^{\pi\left(R_{31} +\frac12 i\sigma \ell_{13}^2\right)\sgn t_{12}t_{23}t_{31}}
\frac{\G\left(i\sigma[R_{23}-R_{12}]-\ell_2\right)}{\G\left(i\sigma[R_{23}-R_{31}]+\ell_3+1\right)}
\\
\int dp\ \frac{\G(i\sigma[R_{23}-R_{31}]+p) \G(\ell_1-\ell_2+p) \G(i\sigma[R_{12}-R_{23}]-\ell_1-p) \prod_\pm \G(-\ell_3^\pm -p)}{\G(-\ell_1-\ell_2-p)} \ .
\end{multline}
This limit has an analog in the case of D-branes in $SU(2)$: the Alekseev--Recknagel--Schomerus limit where maximally symmetric D-branes become fuzzy spheres \cite{ars99}. In the rational $SU(2)$ theory, the algebra of boundary fields on a given D-brane then becomes a finite-dimensional matrix algebra, with the size of the matrices depending on the boundary parameter. In the present $\Hp$ case, the algebra of boundary fields is infinite-dimensional, and may have an interpretation as the algebra of functions on a non-compact, non-commutative $AdS_2$ manifold. The above limit of $\Omega_3$ would then describe the product in this algebra, whose noncommutativity ultimately comes from the lack of worldsheet parity invariance of the $\Hp$ model with boundary. 

\paragraph{Towards the Minkowskian theory.}

Solving the $\Hp$ model may be seen as a step in the study of string theory in the Minkowskian $AdS_3$. On the one hand, this theory is expected to be technically more complicated due to the presence of discrete and spectrally flowed representations in the spectrum \cite{mo00a}, in addition to the purely continuous spectrum of the $\Hp$ model. On the other hand, the formal structure of the theory is probably more conventional, since the symmetry algebra safely factorizes into left- and right-movers. 

Let me explain why the formalism of the present article may be well-suited to studying strings in the Minkowskian $AdS_3$. The conventionality of the formal structure of that theory suggests that $AdS_3$ four-point conformal blocks could be defined using the usual factorization assumption. This assumption is that in the limit $w_{12}\rar 0$, where two fields come close together on the worldsheet, the $s$-channel four-point blocks should factorize into products of three-point blocks: 
\bea
\psset{unit=.25cm}
\pspicture[](-4,-2)(4.5,2)
\psline(3,2)(2,0)(-2,0)(-3,2)
\psline(-3,-2)(-2,0) \psline(3,-2)(2,0)
\rput(3.5,2.2){$2$} \rput(-3.5,2.2){$3$} \rput(-3.5,-2.2){$4$} \rput(3.5,-2.2){$1$}
\rput(0,.6){$s$}
\endpspicture
\underset{w_{12}\rar 0}{\sim} 
\pspicture[](-2,-1.8)(3,1.8)
\psline(-1,-2)(0,0)(-1,2) \psline(0,0)(2,0) \rput(-1.5,2.2){$3$} \rput(-1.5,-2.2){$4$}
\rput(2.4,0){$s$}
\endpspicture
\times
\pspicture[](-3,-1.8)(2,1.8)
\psline(1,-2)(0,0)(1,2) \psline(0,0)(-2,0)
\rput(1.5,2.2){$2$} \rput(1.5,-2.2){$1$}
\rput(-2.4,0){$s$}
\endpspicture
\eea
(It can be seen that the $\Hp$ blocks defined in Section 4.1 do not obey this assumption.) 
Now, this assumption would lead to $s$-channel blocks being singular at $\nu_s\equiv\nu_1+\nu_2=0$, simply because the three-point blocks themselves are. This $\nu_s=0$ singularity takes very characteristic forms when discrete and spectrally flowed representations propagate in the $s$-channel. As was recalled in Section 4.2, an $s$-channel field in a discrete representation would indeed vanish for either $\nu_s<0$ or $\nu_s>0$. I now add that a spectrally flowed field would be a distribution supported at $\nu_s=0$, as can be deduced from \cite{rib05}. Therefore, $\nu$-basis blocks permit an easy characterization of continous, discrete and spectrally flowed $s$-channel modes, based on their behaviour near $\nu_s=0$.

\paragraph{New D-branes in $\Hp$?} 

The relation between the boundary three-point function and the fusing matrix (\ref{chfh}) relies on associating certain boundary fields to the boundary conditions of the model. However, the boundary conditions only have a real parameter $r$, and they are associated a the set of discrete boundary fields, which are far from exhausting the full space of boundary fields ${}_r\Psi_{r'}^\ell$ parametrized by their spin $\ell$ and by $\alpha=r-r'$. Can fusing matrices with generic entries $(\ell,\al)$ be interpreted as three-point structure constants on new maximally symmetric D-branes? If not, why do 
the discrete representations, and only them, give rise to D-branes in $\Hp$?

\appendix

\zeq\section{Some useful formulas}

\subsection{Special functions $\G_b$ and $S_b$}

The special functions $\G_b$ and $S_b$ usually appear in
the study of Liouville theory at parameter $b>0$ and background charge
$Q=b+b^{-1}$. I use the same conventions as \cite{pon03}, where some
more details can be found. The following definitions are valid for
$0<\Re x<Q$:
\bea
\text{log}\G_b(x)&=&\int_{0}^{\infty}\frac{dt}{t}\left\lbrack
\frac{e^{-xt}-e^{-Qt/2}}{(1-e^{-bt})(1-e^{-t/b})}-
\frac{(Q/2-x)^{2}}{2}e^{-t}-\frac{Q/2-x}{t}\right\rbrack \, ,
\label{gammab}
\\
\text{log}S_b&=&\int_{0}^{\infty}\frac{dt}{t}\left\lbrack
\frac{\text{sinh}(\frac{Q}{2}-x)t}
{2\text{sinh}(\frac{bt}{2})\text{sinh}(\frac{t}{2b})}-
\frac{(Q-2x)}{t}\right\rbrack\ .
\label{sb}
\eea
These functions, which are related by $S_b(x)=\frac{\G_b(x)}{\G_b(Q-x)} $, can be extended to meromorphic functions on the complex plane thanks to the shift equations
\bea
\G_b(x+b)= \frac{\sqrt{2\pi}b^{bx-\frac{1}{2}}}{\Gamma(bx)}\G_b(x)  &\scs&
\G_b(x+1/b)=
\frac{\sqrt{2\pi}b^{-\frac{x}{b}+\frac{1}{2}}}{\Gamma(x/b)}\G_b(x)
\\
S_b(x+b) = 2\text{sin}(\pi bx)S_b(x) & \scs & S_b(x+1/b) = 2\text{sin}(\pi
x/b)S_b(x)
\eea
Using the integral representations for the special functions, one can
study their behaviour for $b\rar 0$ while keeping the quantities $x,y$
fixed:
\bea
\Gamma_b(bx)\rar (2\pi b^3)^{\frac12 (x-\frac12)} \Gamma(x)  &\scs&
\Gamma_b(Q-bx)\rar \left(\frac{b}{2\pi}\right)^{\frac12 (x-\frac12)} \ ,
\label{limgb}
\\
 S_b(bx)\rar (2\pi
b^2)^{x-\frac12} \Gamma(x) &\scs& S_b(\frac{1}{2b}+bx)\rar
2^{x-\frac12}\ ,
\label{limsb}
\\
 |\Re y|<\frac12 \Rightarrow S_b(\tfrac{1}{2b}+bx+\tfrac{1}{b}y)&\rar &\left(\frac{\cos \pi y}{2}\right)^{\frac12-x} \exp -\frac{1}{b^2}\int_0^\infty \frac{dt}{t} \left[\frac{\sinh 2yt}{2t\sinh t}-\frac{y}{t}\right]   \ .
\label{limsbr}
\eea

\subsection{Miscellaneous}

The following integral \cite{pst01}, which should be understood as a distribution, appears in eq. (\ref{cgp}).
\bea
\int_\R dy\ e^{i\theta y} |y|^{\al} = \frac{2}{|\theta|^{\al+1}} \G(\al +1) \sin \tfrac{\pi}{2} \al \ .
\label{iyp}
\eea
The following identity, which is valid for three arbitrary real numbers $t_1,t_2,t_3$, is applied to isospins in eqs. (\ref{omt}) and (\ref{otco}). 
\bea
\sgn t_{12} t_{23}t_{31} +\sgn t_{12} + \sgn t_{23} + \sgn t_{31} = 0\ .
\label{signs}
\eea
An integral formula from \cite{gr65} (7.512) is used in eq. (\ref{ifz}):
\bea
\int_0^1dx\ x^{\al-\gamma}(1-x)^{\g-\beta-1}F(\al,\beta,\g,x)=\frac{\G(1+\frac12\al)\G(\g)\G(\al-\g+1)\G(\g-\beta-\frac12\al)}{ \G(\al+1)\G(\frac12\al+1-\beta)\G(\g-\frac12\al)} \ .
\label{izf}
\eea

\subsection{Linear equations for certain Liouville fusing matrices\label{apfl}}

Let me derive linear relations involving the fusing matrices
$F^L_{\eta_1\eta_3}\equiv F^L_{\beta_s\beta_t} \!\!\left[\begin{smallmatrix}\beta_3-\frac{\eta_3}{2b} & \beta_2 \\ \beta_4 &\beta_1-\frac{\eta_1}{2b} \end{smallmatrix}\right]$ and                                          
$F^L_{\eta_2\eta_4}\equiv F^L_{\beta_s\beta_t} \!\!\left[\begin{smallmatrix}\beta_3 & \beta_2 -\frac{\eta_2}{2b}\\ \beta_4 -\frac{\eta_4}{2b}&\beta_1 \end{smallmatrix}\right]$, where $\eta_i=\pm$ are signs. I will use a sequence of Liouville fusing transformations, including some degenerate ones 
whose matrix elements are the $M^{(ij)k}_{\eta\eta'}$ defined in eq. (\ref{mee}):
\bea
\begin{array}{ccccc}
\psset{unit=.35cm}
\pspicture[](-4,-2)(4,2)
\psline(3,2)(2,0)(-2,0)(-3,2)
\psline(-3,-2)(-2,0) \psline(3,-2)(2,0)
\thecoil(-1.6,0)(-1.6,2) \thecoil(1.6,0)(1.6,-2)
\rput(-1.5,-0.6){$\eta$} \rput(1.5,.7){$-\eta$}
\rput(3.5,2.2){$2$} \rput(-3.5,2.2){$3$} \rput(-3.5,-2.2){$4$} \rput(3.5,-2.2){$1$}
\endpspicture
& \overset{M^{(3s)4}_{\eta_3\eta} M^{(1s)2}_{\eta_1,-\eta}}{\longrightarrow} &
\psset{unit=.35cm}
\pspicture[](-4,-1.8)(4,1.8)
\psline(3,2)(2,0)(-2,0)(-3,2)
\psline(-3,-2)(-2,0) \psline(3,-2)(2,0)
\thecoil(-2.2,.4)(-1.6,2.2) \thecoil(2.2,-.4)(1.6,-2.2)
\rput(-2.9,-0.2){$\eta_3$} \rput(2.9,.2){$\eta_1$}
\endpspicture
& \overset{F_{\eta_1\eta_3}^L}{\longrightarrow} & 
\psset{unit=.35cm}
\pspicture[](-3,-3.7)(3,3)
\psline(2,3)(0,2)(0,-2)(2,-3)
\psline(-2,-3)(0,-2)\psline(-2,3)(0,2)
\thecoil(-.4,2.2)(1,3.2) \thecoil(.4,-2.2)(-1,-3.2)
\rput(-.7,1.6){$\eta_3$} \rput(0.7,-1.6){$\eta_1$}
\endpspicture
\\
 = & & & &  M^{(23)t}_{\eta_2\eta_3}\ \downarrow\ M^{(41)t}_{\eta_4\eta_1} 
\\
\psset{unit=.35cm}
\pspicture[](-4,-2)(4,2)
\psline(3,2)(2,0)(-2,0)(-3,2)
\psline(-3,-2)(-2,0) \psline(3,-2)(2,0)
\thecoil(-1.6,0)(-1.6,-2) \thecoil(1.6,0)(1.6,2)
\rput(-1.5,0.6){$\eta$} \rput(1.5,-.7){$-\eta$} 
\endpspicture
& \underset{M^{(4s)3}_{\eta_4\eta} M^{(2s)1}_{\eta_2,-\eta}}{\longrightarrow}& 
\psset{unit=.35cm}
\pspicture[](-4,-1.8)(4,1.8)
\psline(3,2)(2,0)(-2,0)(-3,2)
\psline(-3,-2)(-2,0) \psline(3,-2)(2,0)
\thecoil(2.2,.4)(1.6,2.2) \thecoil(-2.2,-.4)(-1.6,-2.2)
\rput(2.9,-0.2){$\eta_2$} \rput(-2.9,.2){$\eta_4$}
\endpspicture
& \underset{F_{\eta_2\eta_4}^L}{\longrightarrow} & 
\psset{unit=.35cm}
\pspicture[](-3,-3)(3,3.7)
\psline(2,3)(0,2)(0,-2)(2,-3)
\psline(-2,-3)(0,-2)\psline(-2,3)(0,2)
\thecoil(.4,2.2)(-1,3.2) \thecoil(-.4,-2.2)(1,-3.2)
\rput(.7,1.6){$\eta_2$} \rput(-0.7,-1.6){$\eta_4$}
\endpspicture
\end{array}
\eea
 Each choice of $\eta=\pm$ yields a formula for the four matrix elements $F^L_{\eta_2\eta_4}$ in terms of $F^L_{\eta_1\eta_3}$:
 \bea
 \forall \eta=\pm, \qquad F^L_{\eta_2\eta_4} = \sum_{\eta_1,\eta_3} { M}^{(23)t}_{\eta_2\eta_3}{ M}^{(41)t}_{\eta_4\eta_1} F^L_{\eta_1\eta_3} \frac{{ M}^{(3s)4}_{\eta_3,-\eta}}{{ M}^{(4s)3}_{\eta_4,-\eta}} \frac{{ M}^{(1s)2}_{\eta_1\eta}}{{ M}^{(2s)1}_{\eta_2\eta}} \ .
\eea
 Using both choices $\eta=\pm$, one can eliminate $F^L_{\eta_2\eta_4}$ and find the following rank two system of four equations for $F^L_{\eta_1\eta_3}$, where $J\equiv b^{-1}(\beta-\frac{Q}{2})$:
 \begin{multline}
 \forall \eta_2,\eta_4,\qquad \sum_{\eta_1,\eta_3} \frac{\prod_\pm \G(\frac12\pm J_s +\eta_3J_3-\eta_4J_4) \prod_\pm \G(\frac12\pm J_s+\eta_1J_1-\eta_2J_2)}{\prod_\pm \G(\frac12\pm J_t+\eta_3J_3-\eta_2J_2)\prod_\pm \G(\frac12\pm J_t +\eta_1J_1-\eta_4J_4)}
  \\ \times 
  \sin \pi (\eta_2J_2+\eta_3J_3-\eta_1J_1-\eta_4J_4)\ \ F^L_{\eta_1\eta_3} =0\ .
\end{multline}

\acknowledgments{I am grateful to DESY, Hamburg for hospitality and to the Alexander von Humboldt Stiftung for support while part of this work was done. I would like to thank Manfred Herbst, Ingo Runkel and Volker Schomerus for interesting conversations, and Joerg Teschner for collaborating on closely related topics. In addition, I have benefited from helpful comments by Ingo Runkel and Joerg Teschner on the draft of this article.
}

%Bibliography

%\bibliographystyle{morder}
%\bibliography{992}

\end{document}